\documentclass[journal=jacsat,manuscript=article]{achemso}

\setkeys{acs}{articletitle = true}
\setkeys{acs}{chaptertitle = true}
\setkeys{acs}{abbreviations = false}
\SectionNumbersOn 

\usepackage{amsmath}
\usepackage{amssymb}
\usepackage[version=4]{mhchem}

\usepackage{graphicx}
\usepackage{multirow}
\usepackage{siunitx}
\usepackage{xcolor}
\usepackage{braket}
\usepackage{bm}
\usepackage{amsmath}

\usepackage[hidelinks]{hyperref}
\hypersetup{colorlinks=false}

\usepackage{numprint}
\usepackage[utf8]{inputenc}
\usepackage[T1]{fontenc}
\usepackage{mathptmx}
\usepackage[english]{babel}
\usepackage{subfigure}
\usepackage{comment}
\usepackage{siunitx}

\usepackage{xr}
\title{Understanding and mitigating noise in molecular quantum linear response for spectroscopic properties on quantum computers } 

\author{Karl Michael Ziems}
\email{kmizi@kemi.dtu.dk}
\affiliation{Department of Chemistry, Technical University of Denmark, Kemitorvet Building~207, DK-2800 Kongens Lyngby, Denmark.}

\author{Erik Rosendahl Kjellgren}
\affiliation{Department of Physics, Chemistry and Pharmacy, University of Southern Denmark, Campusvej~55, DK-5230 Odense, Denmark.}

\author{Stephan P. A. Sauer}
\affiliation{Department of Chemistry, University of Copenhagen, DK-2100 Copenhagen \O.}
\author{Jacob Kongsted}
\affiliation{Department of Physics, Chemistry and Pharmacy, University of Southern Denmark, Campusvej~55, DK-5230 Odense, Denmark.}
\author{Sonia Coriani}
\affiliation{Department of Chemistry, Technical University of Denmark, Kemitorvet Building~207, DK-2800 Kongens Lyngby, Denmark.}

\date{\today}

\begin{document}

\begin{abstract}
The promise of quantum computing to circumvent the exponential scaling of quantum chemistry has sparked a race to develop chemistry algorithms for quantum architecture. However, most works neglect the quantum-inherent shot noise, let alone the effect of current noisy devices. Here, we present a comprehensive study of quantum linear response (qLR) theory obtaining spectroscopic properties on simulated fault-tolerant quantum computers and present-day near-term quantum hardware. This work introduces novel metrics to analyze and predict the origins of noise in the quantum algorithm, 
proposes an Ansatz-based error mitigation technique, and highlights the significant impact of Pauli saving in reducing measurement costs and noise. 
%
%
Our hardware results using up to cc-pVTZ basis set serve as proof-of-principle for obtaining absorption spectra on quantum hardware in a general approach with the accuracy of classical multi-configurational methods. Importantly, our results exemplify that substantial improvements in hardware error rates and measurement speed are necessary to 
%
%
lift quantum computational chemistry from proof-of-concept to an actual impact in the field.
\end{abstract}

\maketitle

\section{\label{sec:intro}Introduction}

Calculation of many spectroscopic properties of molecules requires
not only the energy of the electronic ground state of the molecular system, but also information about the excited states. In the past decades, several formalisms have been developed to obtain such information without the need to explicitly construct the excited states. Arguably, the two most successful formalisms to this end are the equation of motion (EOM) \cite{rowe_equations-of-motion_1968} and the linear response (LR) \cite{Olsen1985,helgaker_response_2012} formulations. 

Conventionally, to predict the properties of systems with strong correlation, multi-configurational wave function methods such as complete active space (CAS)\cite{Siegbahn1980,Roos1980,Siegbahn1981,Roos:1987} and 
restricted active space (RAS)\cite{Olsen1988,Malmqvist1990} self-consistent field (SCF) have been used in combination with LR.\cite{Olsen1985,Jorgensen:1988,helgaker_response_2012,Helmich-Paris:2019,Delcey:2023}
The drawback of these methods is that the scaling with respect to the system size scales exponentially, with the limit of current technology being 22 electrons in 22 orbitals\cite{Vogiatzis2017} or 26 electrons in 23 orbitals for ``exact'' implementations.\cite{Gao2024}

A promising emerging technology to overcome the exponential scaling of these conventional multi-configurational methods are quantum processing units (QPUs).
At present, fault-tolerant QPUs remain a future aspiration, while the available noisy intermediate-scale quantum (NISQ) devices face significant technological limitations. These limitations not only require extensive error suppression and mitigation strategies but also severely limit the range and quality of executable algorithms. 
For obtaining the molecular ground state wave function, several hybrid quantum-classical algorithms have been proposed, the most prominent examples being the variational quantum eigensolver (VQE)\cite{peruzzo_variational_2014,McClean2016} and approaches based on imaginary- or real-time evolution.\cite{klymko2022real,mcardle2019variational} To reduce the number of qubits, VQE has also been used within the active-space approximation.\cite{Mizukami2020,Takeshita2020,Sokolov2020,Bierman2023}

To utilize the wave function on a quantum device for property simulations, the EOM framework was extended to quantum EOM (qEOM).\cite{McClean2017, Ollitrault2020}
In the context of variational formulations, the EOM and LR formulations are identical for complete space unitary coupled-cluster,\cite{Taube2006} and very similar for other Ans\"atze and in reduced spaces, hence qEOM and qLR were developed in parallel and are combined in the following.
The qEOM formalism has been extended to many different use cases including
multi-component-EOM,\cite{Pavoevi2021a} QED-EOM,\cite{Pavoevi2021} spin-flip qEOM\cite{Pavoevi2023}, qEOM for thermal averages of quantum states,\cite{https://doi.org/10.48550/arxiv.2406.04475} and qEOM for non-adiabatic molecular dynamics.\cite{Gandon2024} 
The original qEOM formulation has also seen improvements to its parameterizations as self-consistent (sc-qEOM/LR)\cite{Asthana2023,Kumar2023} and projected (proj-qEOM/LR) variant.\cite{Kumar2023} Additionally, LR has been implemented by solving auxiliary response quantum states via VQE.\cite{cai2020quantum,huang2022variational}
A crucial step to near-term application, qEOM/LR were recently formulated within the active-space approximation with orbital-optimization (oo) as a powerful hybrid-classical algorithm opening for simulations beyond minimal basis on near-term devices. Apart from the naive implementation oo-qEOM\cite{Jensen2024} and oo-qLR\cite{Ziems2024}, eight different parametrizations were introduced by Ziems et al. using combinations of naive, self-consistent, state-transfer, and projected formalisms.\cite{Ziems2024} The most suitable NISQ-era algorithms were found to be oo-qLR, oo-proj-qLR, and oo-allproj-qLR. 
Moreover, algorithmic improvements to the qEOM/LR formalism such as the Davidson solver sc-qEOM/LR\cite{Kim2023,Reinholdt2024} and
a reduced density matrix formulation of oo-qLR\cite{Von_Buchwald2024-pp} have recently 
been reported.

So far, the work in this field has mostly focused on 
algorithmic developments and testing on noise-free simulators that ignored not just the device noise of NISQ-QPUs but also the unavoidable quantum mechanical shot/sampling noise of fault-tolerant quantum computers. This work addresses these shortcomings and reports qLR results not only on a simulated fault-tolerant quantum computer (i.e., a shot-noise simulator) but also on currently available NISQ quantum hardware. Moreover, we analyze in detail how the quantum mechanical nature of quantum devices impacts the qLR/EOM algorithm and compare algorithms, systems, and excited state results using various (novel) metrics. For the mitigation of noise, we introduce our own Ansatz-based read-out and error mitigation and on-the-fly Pauli saving.

Importantly, we provide a proof-of-principle that absorption spectra can be obtained on a real quantum device using qLR/EOM formulations with a triple-zeta basis set. For this, we choose a small molecule, yet
with a general and accurate ansatz that can reproduce classic quantum chemistry. Thus, this study provides insight into the maturity of chemistry on quantum computers not from the viewpoint of a technical quantum advantage but of the possibilities and limitations for a chemist to obtain useful results on QPUs in the future. 

\section{\label{sec:theo}Theory}

Throughout the paper, we use $p,q,r,s$ as general (spatial) orbital indices, $a,b,c,d$ as virtual indices, $i,j,k,l$ as inactive indices, and $v,w,x,y$ as active indices.
The notation $v_a$ indicates an active index that is virtual in the Hartree-Fock reference, while $v_i$ stands for an active index that is inactive in the Hartree-Fock reference.

\subsection{\label{sec:theo:AS} Active space wave function} 

To reduce quantum computational costs and leverage the complementing strength of classic and quantum architecture at best, we employ an active space wave function with orbital optimization. The wave function is split according to 
\begin{equation}
        \left|0\left(\theta\right)\right> = \left|I\right>\otimes \left|A\left(\theta\right)\right> \otimes \left|V\right>~,
\end{equation}
where $\ket{I}$, $\ket{A(\theta)}$, and $\ket{V}$ are the inactive, active, and virtual parts, respectively. The active part is prepared on a (simulated) quantum device using a parameterized unitary transformation (Ansatz)
\begin{equation}
    \ket{A(\theta)} = U(\theta)\ket{A}
\end{equation}
and encompasses orbitals that are expected to contribute largely to electron correlation. Every operator is decomposed into the three spaces as well
\begin{equation}
    \hat{O} = \hat{O}_I\otimes \hat{O}_A\otimes \hat{O}_V,
\end{equation}
which yields the expectation value as
\begin{equation}
\left<0\left(\boldsymbol{\theta}\right)\left|\hat{O}\right|0\left(\boldsymbol{\theta}\right)\right> = \left<I\left|\hat{O}_{I}\right|I\right>\otimes \left<A\left(\boldsymbol{\theta}\right)\left|\hat{O}_{A}\right|A\left(\boldsymbol{\theta}\right)\right> \otimes\left<V\left|\hat{O}_{V}\right|V\right>~. \label{eq:op_splitting}
\end{equation}
Here, the inactive and virtual parts are trivial to calculate on a classical computer, while the multi-configurational active part is calculated on (simulated) quantum hardware.

The active space unitary, $U(\theta)$, can utilize any common quantum computing wave function Ansatz such as k-UpCCGSD,\cite{Lee2018} QNP,\cite{Anselmetti2021} tUPS,\cite{Burton2024} and variants of ADAPT.\cite{Grimsley2019,Tang2021,Feniou2023,Burton2023,Majland2023,Anastasiou2024} In this work, we use trotterized unitary coupled-cluster with singles and doubles (tUCCSD)
\begin{equation}
    \ket{\text{tUCCSD}\left(\theta\right)} = \prod_k^{N_\text{SD}}\boldsymbol{U}_{k}\left(\theta\right)\ket{A}
\end{equation}
where,
\begin{equation}
    \boldsymbol{U}_{k}\left(\theta\right) = \prod_l^{N_\text{Pauli}} \text{e}^{i\theta_k\hat{P}_{k,l}}
\end{equation}
with $\hat{P}_{k,l}$ being the $l$'th Pauli string resulting from the mapping of the $k$'th fermionic operator to Pauli strings. The index $k$ runs over all $N_\mathrm{SD}$ number of singles and doubles excitation, and the index $l$ over all $N_\mathrm{Pauli}$ Pauli strings for a given $k$. 
For the tUCCSD ansatz, the fermionic operators are constructed as 
$$\theta_{v_I}^{v_A}\left(\hat{a}^\dagger_{v_A} \hat{a}_{v_I} - \hat{a}^\dagger_{v_I}\hat{a}_{v_A}\right)$$ 
for the single excitations and 
$$\theta_{v_I v_J}^{v_A v_B}\left(\hat{a}^\dagger_{v_A}\hat{a}^\dagger_{v_B}\hat{a}_{v_J}\hat{a}_{v_I} - \hat{a}^\dagger_{v_I}\hat{a}^\dagger_{v_J}\hat{a}_{v_B} \hat{a}_{v_A}\right)~,$$ with the constraints $v_I<v_J$ and $v_A<v_B$, for the double excitations (capitalized indices here refer to spin orbitals).

The orbital optimization is included via non-redundant rotations of the type \textit{inactive to active}, \textit{inactive to virtual}, and \textit{active to virtual},
$pq\in\left\{vi,ai,av\right\}$.
The orbital rotation parameters are 
\begin{equation}
    \hat{\kappa}\left(\boldsymbol{\kappa}\right) = \sum_{p>q}\kappa_{pq}\hat{E}^-_{pq}
\end{equation}
with $\hat{E}^-_{pq} = \hat{E}_{pq} - \hat{E}_{qp}$ and the singlet single-excitation operator $\hat{E}_{pq} = \hat{a}_{p,\alpha}^\dagger\hat{a}_{q,\alpha} + \hat{a}_{p,\beta}^\dagger\hat{a}_{q,\beta}$. This formally results in the oo-tUCCSD wave function
\begin{equation}
    \left|\text{oo-tUCCSD}\right> = \mathrm{e}^{-\hat{\kappa}\left(\boldsymbol{\kappa}\right)}\prod_k^{N_\text{SD}}\boldsymbol{U}_{k}\left(\theta\right)\left|\text{CSF}\right>, 
\end{equation}
where $\left|\text{CSF}\right>$ is a single configuration state function reference.
However, instead of acting on the state vector, the orbital rotations are used to transform the integrals\cite{Helgaker2000}
\begin{align}
h_{pq}\left(\boldsymbol{\kappa}\right) &= \sum_{p'q'} \left[\mathrm{e}^{\boldsymbol{\kappa}}\right]_{q'q}h_{p'q'}\left[\mathrm{e}^{-\boldsymbol{\kappa}}\right]_{p'p}\\
g_{pqrs}\left(\boldsymbol{\kappa}\right) &= \sum_{p'q'r's'}\left[\mathrm{e}^{\boldsymbol{\kappa}}\right]_{s's}\left[\mathrm{e}^{\boldsymbol{\kappa}}\right]_{q'q}g_{p'q'r's'}\left[\mathrm{e}^{-\boldsymbol{\kappa}}\right]_{p'p}\left[\mathrm{e}^{-\boldsymbol{\kappa}}\right]_{r'r},
\end{align}
of the Hamiltonian
\begin{equation}
    \hat{H}\left(\boldsymbol{\kappa}\right) = \sum_{pq}h_{pq}\left(\boldsymbol{\kappa}\right)\hat{E}_{pq} + \frac{1}{2}\sum_{pqrs}g_{pqrs}\left(\boldsymbol{\kappa}\right)\hat{e}_{pqrs}
\end{equation}
where $\hat{e}_{pqrs} = \hat{E}_{pq}\hat{E}_{rs} - \delta_{qr}\hat{E}_{ps}$ is the two-electron singlet excitation operator. 

The ground state wave function and its energy can now be found by variational minimization of the parameters $\boldsymbol{\theta}$ and $\boldsymbol{\kappa}$,
\begin{equation}
    E_\text{gs} = \min_{\boldsymbol{\theta}, \boldsymbol{\kappa}}\left<\text{oo-tUCCSD}\left(\boldsymbol{\theta}\right)\left|\hat{H}\left(\boldsymbol{\kappa}\right)\right|\text{oo-tUCCSD}\left(\boldsymbol{\theta}\right)\right>.
\end{equation}
In the context of quantum computing, this minimization is known as the orbital-optimized variational quantum eigensolver (oo-VQE) algorithm.\cite{Mizukami2020,Takeshita2020,Sokolov2020,Bierman2023}

\subsection{\label{sec:theo:qLR} Quantum Linear Response} 

For sake of conciseness, we refer to the original quantum work\cite{Ziems2024} or classic linear response literature\cite{helgaker_response_2012} for detailed information and only introduce the most important equations here. 

The qLR framework allows us to obtain excited state energies and properties within first-order time-dependent perturbation theory on top of any variationally obtained ground state wave function. Specifically, 
the generalized eigenvalue problem
\begin{align}
  \textbf{E}^{[2]}  \boldsymbol{\beta}_k =  \omega_k \textbf{S}^{[2]}\boldsymbol{\beta}_k  
\label{eq:LR_nopert}
\end{align}
gives excitation energies $\omega_k$ and corresponding excitation vectors $\boldsymbol{\beta}_k$. Therein, we define a Hessian and metric matrix
\begin{align}
 \textbf{E}^{[2]} &= \begin{pmatrix}
    \boldsymbol{A} & \boldsymbol{B}          \\[0.3em]
      \boldsymbol{B}^* & \boldsymbol{A}^*           
     \end{pmatrix}, \quad 
     \textbf{S}^{[2]} = \begin{pmatrix}
    \boldsymbol{\Sigma} & \boldsymbol{\Delta}          \\[0.3em]
     -\boldsymbol{\Delta} ^* &  -\boldsymbol{\Sigma}^*           
     \end{pmatrix}, \quad
\end{align}
with the submatrices
\begin{align}
    \boldsymbol{A} &= \boldsymbol{A}^\dagger,\quad A_{IJ} =  \left<0\left|\left[\hat{X}_{I}^\dagger,\left[\hat{H},\hat{X}_{J}\right]\right]\right|0\right>\label{eq:A}\\
    \boldsymbol{B} &= \boldsymbol{B}^\mathrm{T},\quad B_{IJ} =  \left<0\left|\left[\hat{X}_{I}^\dagger,\left[\hat{H},\hat{X}_{J}^\dagger\right]\right]\right|0\right>\label{eq:B}\\
    \boldsymbol{\Sigma} &= \boldsymbol{\Sigma}^\dagger,\quad \Sigma_{IJ} = \left<0\left|\left[\hat{X}_{I}^\dagger,\hat{X}_{J}\right]\right|0\right>\label{eq:Sigma}\\
    \boldsymbol{\Delta} &= -\boldsymbol{\Delta}^\mathrm{T},\quad \Delta_{IJ} = \left<0\left|\left[\hat{X}_{I}^\dagger,\hat{X}_{J}^\dagger\right]\right|0\right>. \label{eq:Delta}
\end{align}
Here, $\hat{X}_l \in \left\{ \hat{Q}_\mu, \hat{R}_n \right\}$, with $\hat{Q}_\mu$ being a generic orbital rotation operator and $\hat{R}_n$ a generic active space excitation operator, exploiting, similar to what is done for the ground state wave function, an active space approach. 
Ziems et al.~\cite{Ziems2024} introduced eight different parameterizations to $\hat{Q}_\mu$ and $\hat{R}_n$ and truncated the latter to the level of singles and doubles. Out of these, three were deemed near-term suitable, namely
\begin{itemize}
\item Naive LR (\textit{naive LR}),
 using $\hat{R}=\hat{G}$ and $\hat{Q}=\hat{q}$; 
\item Projected LR (\textit{proj LR}), using $\hat{R}=\hat{G}\left|0\right>\left<0\right| - \left<0\left|\hat{G}\right|0\right>$ and $\hat{Q}=\hat{q}$; 
\item All projected LR (\textit{allproj LR}) using $\hat{R}=\hat{G}\left|0\right>\left<0\right| - \left<0\left|\hat{G}\right|0\right>$ and $\hat{Q}=\hat{q}\left|0\right>\left<0\right|$; 
\end{itemize}
with the naive orbital rotation operator 
\begin{equation}
    \hat{q}_{pq} = \frac{1}{\sqrt{2}}\hat{E}_{pq}
\end{equation}
and the naive active-space spin-adapted singlet single and double excitation operator\cite{Paldus1977,Piecuch1989,jod15,Packer1996}
\begin{align}
    \hat{G} \in \Bigg\{\frac{1}{\sqrt{2}}\hat{E}_{v_av_i},\quad &\frac{1}{2\sqrt{\left(1+\delta_{v_av_b}\right)\left(1+\delta_{v_iv_j}\right)}}\left(\hat{E}_{v_av_i}\hat{E}_{v_bv_j} + \hat{E}_{v_av_j}\hat{E}_{v_bv_i}\right),\\\nonumber
    &\frac{1}{2\sqrt{3}}\left(\hat{E}_{v_av_i}\hat{E}_{v_bv_j} - \hat{E}_{v_av_j}\hat{E}_{v_bv_i}\right)\Bigg\}.
\end{align}
The spin-adapted operators guarantee that only singlet excitations are calculated in the qLR. For the specific case of full space simulation, i.e.\, when the active space spans the full space, no orbital rotations are present, and \textit{proj qLR} and \textit{allproj qLR} become the same method. 
For all methods, $\boldsymbol{\Delta}=\boldsymbol{0}$.

While \autoref{eq:LR_nopert} is diagonalized classically, the matrix elements \autoref{eq:A}-\ref{eq:Delta} are obtained as expectation value measurements of the ground state wave function using the active space separation in the wave function and qLR operators introduced above. 

Within this formalism, oscillator strengths for a given excited state $k$ are obtained as
\begin{align}
    f_k &=  \frac{2}{3} \omega_k \sum_\gamma \left| \braket{0| [\hat{\mu}_\gamma,\hat{\tilde{O}}_k] |0}\right|^2. \label{eq:osc_strength}
\end{align}
with the normalized excitation operator 
\begin{align}
    \hat{\tilde{O}}_k &= \frac{\hat{O}_k}{\sqrt{\braket{k|k}}} 
    = \frac{\hat{O}_k}{\sqrt{\braket{ 0|[ \hat{O}_k, \hat{O}_k^\dagger ] |0}}}\\
    \hat{O}_k &= \sum_{l \in \mu,n}\left({Z}_{k,l} \hat{X}_l^\dagger + Y_{k,l}\hat{X}_l\right), \label{eq:exc_op}
\end{align}
and $Z_{k,l}$ and $Y_{k,l}$ being the weights in $\boldsymbol{\beta}_k$ of the corresponding qLR operator $\hat{X}_l$. In singles and doubles excitation, qLR represents a up to 6-RDM property for the \textit{proj qLR} and \textit{allproj qLR}, and a up to 4-RDM property for the \textit{naive qLR}.\cite{Von_Buchwald2024-pp}

\subsection{\label{sec:theo:workflow} Quantum Computational Workflow} 

Having briefly recapitulated the wave function ansatz and qLR parametrizations of interest, we now describe the methods we implemented in our in-house quantum computational software  \texttt{SlowQuant}\cite{slowquant} to understand and mitigate quantum errors in the qLR formalism and reduce computational costs.

\subsubsection{\label{sec:theo:PS} On-the-fly Pauli saving}

As the active space part of each expectation value (cf. \autoref{eq:op_splitting}) is evaluated on quantum architecture, each fermionic operator therein is expressed as a sum of Pauli strings to be measured on the wave function:
\begin{align}
    \braket{A|\hat{O}_A|A}  = \sum_{vw\cdots} h_{vw\cdots} \braket{A| a_v^\dagger a_w \cdots |A}  = \sum_l c_l \braket{A|  P_l |A}. \label{eq:exp_pauli}
\end{align}
The chosen mapping determines the exact Pauli string decomposition, and each Pauli string $P_l$ has a corresponding coefficient $c_l$ connected to the fermionic integral $h_{pq\cdots}$ and to the mapping procedure. 

During the quantum computational workflow, from wave function optimization to molecular property, many expectation values are evaluated in order to assess energies, gradients, and second derivatives/hessians. These will have common Pauli strings. Thus, on an on-the-fly basis, every new Pauli string distribution is saved and stored in memory. This storage is accessed and updated during the computational workflow as new Pauli strings are needed. Importantly, only the most connected clique within qubit-wise commutation is stored and this only for a given parametrized circuit. 

At its core, this approach means that a given Pauli string always has the same variance regardless of in which expectation value or where during the algorithm it appears. This is also true for measurement methods based on quantum tomography\cite{d2003quantum} and approximations thereof, like classic shadows\cite{huang2020predicting} and matrix product state tomography.\cite{cramer2010efficient} This implies that our findings apply to a wide range of (reduced and approximated) measurement schemes. 

\subsubsection{\label{sec:theo:MA0} 
Ansatz-based read-out and gate error mitigation}
The error mitigation in this work is an extension of read-out error mitigation (REM)\cite{maciejewski2020mitigation,bravyi2021mitigating} based on calculating a confusion matrix, $\boldsymbol{M}_\text{REM}$,
\begin{equation}
    \boldsymbol{p}_\text{mitigated} = \boldsymbol{M}_\text{REM}^{-1}\boldsymbol{p}_\text{raw}
\end{equation}
with $\boldsymbol{p}_\text{raw}$ being the bit-string probability vector, and $\boldsymbol{p}_\text{mitigated}$ being the error mitigated bit-string quasi-probability vector.
The elements of the confusion matrix are calculated as the probability (Pr) of measuring bit-string $b_j$ when the circuit is prepared to produce bit-string $b_i$,
\begin{equation}
    M_{\text{REM},ij} = \mathrm{Pr}\left(\left.b_j\right|b_i\right)
\end{equation}

Instead of only encoding read-out error, both read-out error and gate error can be encoded into the confusion matrix.
This is achieved by including the ansatz when measuring the bit-strings, $b_j$.
To know what the prepared bit-string $b_i$ is, all the parameters in the ansatz are set to zero.
This makes the circuits Clifford-circuits, thus this error mitigation can also be seen as a form of Clifford data regression.\cite{Czarnik2021}
The circuit preparation to construct $b_i$ takes the form,
\begin{equation}
    \text{circ}\left(b_i\right) = \boldsymbol{U}\left(\boldsymbol{0}\right)\boldsymbol{X}_i
\end{equation}
with $\boldsymbol{X}_i$ being X-gates placed such that $b_i$ is created.
The final form of the error-mitigation is now
\begin{equation}
    \boldsymbol{p}_\text{mitigated} = \boldsymbol{M}_{\boldsymbol{U}_0}^{-1}\boldsymbol{p}_\text{raw}~.
\end{equation}

It should be noted that, in the form currently used in this work, this error mitigation technique is exponentially scaling in the number of required measurements.

\subsubsection{\label{sec:theo:post}Post-processing}

To post-process noisy result runs we use the eigenvalues of the Hessian, $\textbf{E}^{[2]}$. 
If we obtain a negative eigenvalue, the simulation is disregarded as non-physical. 
From a classical point of view, the reason for this is that negative eigenvalues mean that we are not in a wave function minimum. This is relevant for a scenario where we also perform VQE on the device.
In the case of qLR on top of a wave function with confirmed reached minimum, a negative eigenvalue corresponds to a ``lost'' excitation due to noise in the qLR formalism. In other words, an excitation has become a de-excitation.

\subsubsection{\label{sec:theo:metrics} Metrics for quantum error}
Each quantum computer, either near-term or fault-tolerant, observes shot noise due to the stochastic nature of quantum mechanics. Thus, many measurements (shots) for each Pauli string are necessary to obtain the final result as its mean. We here introduce metrics to understand 
the impact of this 
for the qLR algorithms.

The expectation value of an operator $\hat{O}_A$ is obtained via measurements of Pauli strings on a wave function (cf. \autoref{eq:exp_pauli}). Each measurement yields a bit-string distribution from which the expectation value is obtained as the sum of the means for each Pauli string distribution, i.e.\ \mbox{$\mu_{\hat{O}_A} = \sum_l \mu_{P_l}$}.
Additionally, the standard deviation 
of an operator's expectation value, $\sigma_{\hat{O}_A}$ can be obtained from the distributions: 
\begin{align}
    \sigma_{\hat{O}_A} &= \sqrt{\sum_l \sigma^2_{P_l}} \label{eq:qstd}\\
    \sigma^2_{P_l} &=4 \Re\{c_l^2\} (p_1-p_1^2) \label{eq:qstd_2}
\end{align}
with $p_1$ being the probability of measuring $1$ for a given Pauli string, and $\sigma_{P_l}$ being the standard deviation of a single Pauli string measurement.
Only the real part of the coefficients $c_l^2$ is considered since the final expectation values of interest are real.
Hence, the variance in the imaginary part does not contribute. Performing this analysis on an ideal simulator (shot-noise free) gives the quantum-based standard deviation in the limit of infinite many shots. 
This procedure is also at the basis of shot balancing.\cite{Zhu2024,Crawford2021,Wecker2015,https://doi.org/10.48550/arxiv.2004.06252}

To analyze the noise in our qLR formalism we use various standard deviations. 

First, we sample the qLR excitation energies for a given number of shots per Pauli string by performing many individual qLR simulations on a shot noise simulator, which provides the \textit{sampled} standard deviation, 
$\sigma_k$, for each state $k$. This is the most accurate 
(yet expensive) and ``black-box'' insight into how each excited state is influenced by the quantum mechanical nature of quantum computers. 

Second, we construct the qLR matrices, \autoref{eq:A}-\ref{eq:Sigma}, with the standard deviation for each matrix element 
obtained via \autoref{eq:qstd}. Then, each matrix is analyzed in terms of their average standard deviation, 
$\overline{\sigma}_M$ with $\boldsymbol{M}\in{\boldsymbol{A},\boldsymbol{B},\boldsymbol{\Sigma}}$, 
which we will refer to as 
\textit{matrix} std from here on. This tells us how the noise spreads through the various matrices of qLR and allows us to compare between different molecules and qLR parametrizations. Additionally, we construct a matrix standard deviation $\overline{\sigma}_{M,nc}$ where we set the coefficient in \autoref{eq:qstd_2} to $c_l = 1$ for all $P_l$. Comparing this to $\overline{\sigma}_M$ allows us to assess whether the standard deviation is mainly driven by the deviation in the Pauli string or by large coefficients.

Third, we use the procedure above to obtain the average standard deviation in each row, $\overline{\sigma}_M(\hat{X}_l)$ of each matrix $\boldsymbol{M}$. This corresponds to the standard deviation associated with each qLR operator $\hat{X}_l$ in matrix $\boldsymbol{M}$. Combining each operator's standard deviation with the contribution of each operator to a given excited state (extracted from the excitation vector, $\boldsymbol{\beta}_k$) allows us to obtain an expected standard deviation in each excited state from the qLR formalism, 
\begin{align}
    \overline{\sigma}_{M,k} = \sum_l \overline{\sigma}_M(\hat{X}_l) |\beta_{k,l}|^2~,
\end{align}
which is named \textit{state-specific} standard deviation.
Note that this is an analysis routinely done in classic quantum chemistry to understand the contributions of operators (and hence orbital transitions) to an excited state, but it is here combined with the deviation in quantum operators to understand the impact of quantum noise. 

Additionally, the condition number and the average {\em coefficient of variation} (CV) of each matrix $\boldsymbol{M}$ are calculated as indicators of the matrix's susceptibility to noise.\cite{kjellgren2024divergences} The CV is defined as the ratio of the standard deviation (calculated as in \autoref{eq:qstd}) and the mean. This is performed for each matrix element and averaged over.

Since device noise scales with shot noise, the information from this in-depth analysis will also help understand where algorithmic bottlenecks arise in noisy near-term machines. We note that such analysis can be performed for any quantum algorithm based on expectation value constructions. 

\section{\label{sec:comp} Computational Details}

All simulations were performed using our in-house quantum computational chemistry software \texttt{SlowQuant},\cite{slowquant} where we implemented all methods and analysis tools described above, along with
interfaces to \texttt{Qiskit},\cite{qiskit2024} IBM Quantum, and to \texttt{PySCF}\cite{Sun2015,Sun2017,Sun2020} for the 
integrals.

On the shot noise simulator, we studied qLR of H$_2$, LiH(2,2), H$_4$ and BeH$_2$(4,4) in a STO-3G\cite{hehre1969a,hehre1970a} basis set, while we run LiH(2,2) in both the STO-3G and cc-pVTZ\cite{dunning1989a,prascher2011a} basis sets on IBM Osaka. Here, a CAS-like active-space notation WF($n$,$o$) was adopted, where $n$ is the number of electrons in the active space and $o$ is the number of spatial orbitals in the active space.

We used an oo-tUCCSD Ansatz for the wave function, qLR at the singles and doubles level (qLRSD), and Parity mapping for the fermionic-to-qubit mapping. Qubit-wise-commutivity (QWC)\cite{Gokhale2020} was employed in this work. The QWC algorithm used to find an approximate solution to the minimum clique coverage problem is a first-fit algorithm. For all simulations, the oo-VQE was performed classically with a shot-free simulator. The findings of this work, which focuses on qLR, are not reliant on having used oo-VQE for the ground state wave function optimization.

In the shot noise simulations we used a sampling of many qLR runs to obtain the sampled std. For this, 100,000 shots per Pauli string were used and 1000 individual runs for the sampling of H$_2$ and LiH(2,2) and 250 individual runs for H$_4$ and BeH$_2$(4,4). 

For the hardware simulations, we used the same specifications as for the shot noise simulations, apart from an increase of shots per Pauli to 500k (and 1M) representing a total shot budget of 4.5 million (9 million) for each qLR run and 2 million (4 million) for the error mitigation.

\section{\label{sec:results} Results and Discussion}

\subsection{Simulated quantum device and noise analysis}
We first present results from runs on simulated fault-tolerant quantum computers, focusing on analyzing and understanding the error induced by the fundamental quantum nature of quantum computers (shot/sampling noise) on qLR. For qLR results on noise-free quantum computers (infinite shots limit) we refer to the literature.\cite{Ziems2024,Reinholdt2024,Von_Buchwald2024-pp}

\subsubsection{\label{sec:results:PS}Pauli saving impact}

\autoref{tab:PS} shows the amount of Pauli Strings that need to be measured for naive, proj, and allproj qLR for the four systems of interest, namely H$_2$, H$_4$, LiH(2,2), and BeH$_2$(4,4). It differentiates the amounts based on the introduced saving procedures of Pauli saving (PS, see \autoref{sec:theo:PS}) and qubit-wise commutivity (QWC) against using the simple approach of evaluating each expectation value independently (referred to as ``none''). 

Using PS and QWC together leads to a dramatic reduction of the number of 
Pauli strings to be measured, by 75-99\% or up to two orders of magnitude. This saving 
increases with system (and thus qubit) size and will become even more advantageous for larger systems than the ones studied here. For the full space simulations, i.e.,  H$_2$ and H$_4$, proj qLR requires slightly more measurements than naive qLR comparing PS+QWC and none. This is expected as the proj qLR inserts projections onto the wave function that leads to more matrix elements and in turn Pauli strings to be evaluated (see working equations Ref \cite{Ziems2024}). On the other hand, the active space simulations, i.e.\ LiH(2,2) and BeH$_2$(4,4), have the opposite trend with allproj qLR and proj qLR needing significantly less measurements. This is again in line with the qLR working equations as naive, proj and allproj consist of 14, 10 and 7 generic matrix element terms, respectively, where the saving in terms comes from the orbital rotation parts that are not present in full space systems. Less generic matrix element terms leads to less Pauli string evaluations. 

\begin{table}[htbp]
\caption{Number of Pauli strings to be evaluated using  Pauli saving (PS) and qubit-wise commutivity (QWC) for different molecules in minimal basis, for different near-term qLR implementations.}
\begin{center}
\begin{tabular}{l|ccc}
\multicolumn{1}{c|}{\textbf{}} & \multicolumn{1}{c|}{\textbf{PS + QWC}} & \multicolumn{1}{c|}{\textbf{QWC}} & \multicolumn{1}{c}{\textbf{none}} \\ \hline
 & \multicolumn{ 3}{c}{\textbf{H}$\boldsymbol{_2}$} \\ \hline
\textbf{naive} & 9 & 35 & 42 \\ 
\textbf{proj} & 9 & 38 & 64 \\  
 \hline
 & \multicolumn{ 3}{c}{\textbf{LiH(2,2)}} \\ \hline
\textbf{naive} & 9 & 1,118 & 1,774 \\ 
\textbf{proj} & 9 & 922 & 1,491 \\ 
\textbf{allproj} & 9 & 447 & 715 \\ 
 \hline
 & \multicolumn{ 3}{c}{\textbf{H}$\boldsymbol{_4}$} \\ \hline
\textbf{naive} & 699 & 22,908 & 76,406 \\ 
\textbf{proj} & 774 & 15,941 & 83,041 \\ 
 \hline
 & \multicolumn{ 3}{c}{\textbf{BeH}$\boldsymbol{_2}$\textbf{(4,4)}} \\ \hline
\textbf{naive} & 822 & 104,096 & 420,132 \\ 
\textbf{proj} & 753 & 64,076 & 309,531 \\ 
\textbf{allproj} & 753 & 44,510 & 227,781 \\  
\end{tabular}
\end{center}
\label{tab:PS}
\end{table}

For brevity, we only show in the following the simulation results of LiH(2,2) and H$_4$, and refer the reader to the SI for details on H$_2$ and BeH$_2$(4,4). These show mostly the same behaviour and at times similarities or differences will be highlighted in the text. 

\begin{figure}[htbp]
    \centering 
    \includegraphics[width=.75\textwidth]{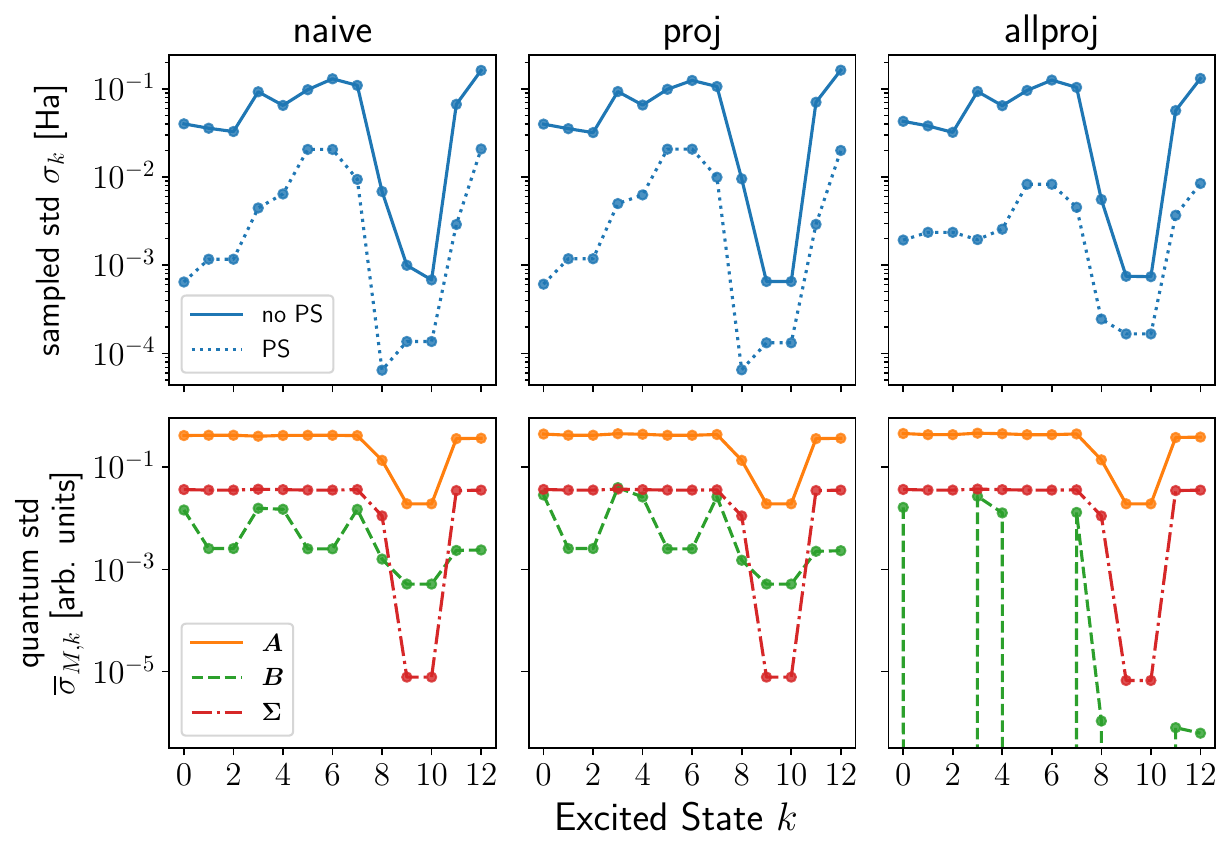}
    \caption{Standard deviation (std) analysis of naive (left), proj (middle) and allproj (right) qLR of LiH. The top row shows the samples std, $\sigma_k$, with (dotted line) and without (solid line) Pauli saving (PS). The bottom row shows the state-speciifc std, $\overline{\sigma}_{M,k}$ for the qLR matrices $\boldsymbol{M}=\boldsymbol{A}$ (orange solid), $\boldsymbol{M}=\boldsymbol{B}$ (green dashed) and $\boldsymbol{M}=\boldsymbol{\Sigma}$ (red dashed-dotted). Recall \autoref{sec:theo:metrics} and see SI \autoref{SI:sec:add_shotnoise} for more details on the noise metrics and additional figures, respectively.}
    \label{fig:LiH_all}
\end{figure}

\begin{figure}[htbp]
    \centering 
    \includegraphics[width=.5\textwidth]{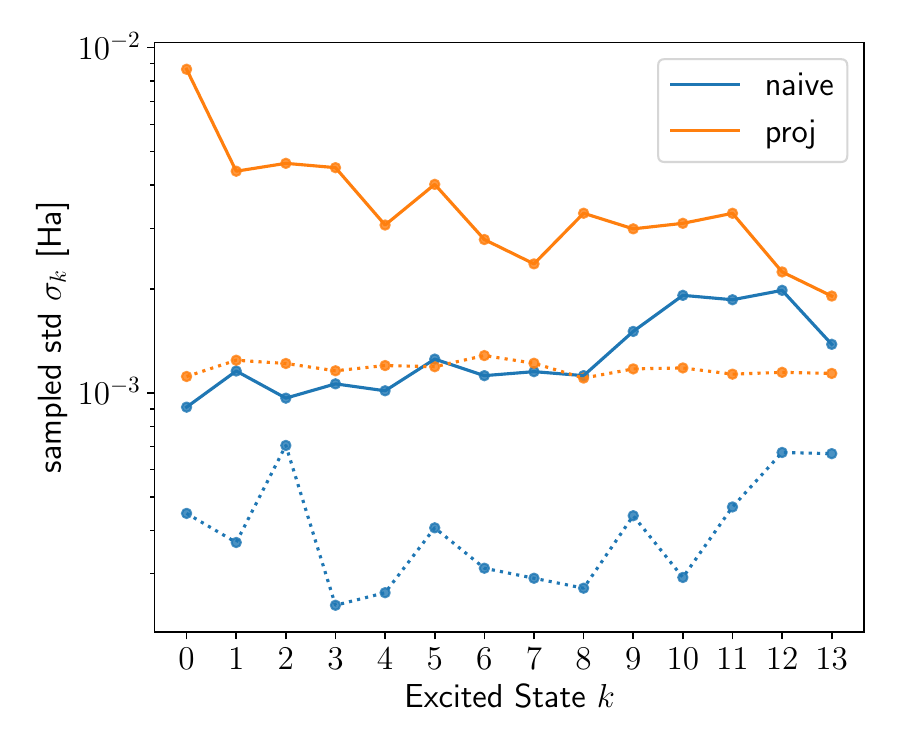}
    \caption{Sampled std $\sigma_k$ with (dotted line) and without (solid line) Pauli saving for H$_4$ naive (blue) and proj (orange) qLR.}
    \label{fig:H4_save_or_not}
\end{figure}

The sampled stds, $\sigma_k$, from shot noise simulation in \autoref{fig:LiH_all} (top row or SI \autoref{SI:fig:LiH_save_or_not}) and \autoref{fig:H4_save_or_not} show that Pauli saving leads to a reduction in shot-noise-induced deviation of up to a factor of 100 for selected excited states (solid vs. dotted lines). Interestingly, this effect is more pronounced for active space simulations.
Even more dramatically, 13/11/9\% of the shot noise runs of LiH(2,2) naive/proj/allproj qLR failed without PS, as they resulted in a negative eigenvalue from the Hessian and were removed in post-processing as non-physical results (cf. \autoref{sec:theo:post}).
For BeH$_2$(4,4), this increased to $>90\%$ without PS. This happening on shot noise simulator, i.e.\ a fault-tolerant quantum computer, for 100k shots per Pauli strings\footnote{Note that we chose 100k shots as realistic number because this was the upper limit of shots per circuit on IBM hardware at the time.} for such small systems, shows how advances in measurement number and speed will be crucial for any useful application of quantum computing in chemistry in the future. We showed that schemes like PS can assist alongside hardware improvements, in this specific case reducing both the number of shots and noise-induced std by up to two orders of magnitude.

The reason behind the noise reduction obtained using PS is that it preserves symmetries even for noisy results since the same Pauli string has always the same noise. In SI \autoref{SI:sec:PS}, we provide a more mathematical reasoning for the example of a 2x2 matrix. Importantly, this finding holds for any hybrid quantum computing method that solves a (generalized) eigenvalue equation and might extend to other algorithms. 

\subsubsection{\label{sec:results:noise}Noise analysis}

In \autoref{fig:LiH_all} for LiH (and SI \autoref{SI:sec:add_shotnoise} for other systems), we see that the sampled std, $\sigma_k$, is different for each excited state. This means that in our qLR approach the inherent quantum mechanical shot noise impacts the excited states differently. Moreover, the trend and overall std can be different in the three near-term qLR formulations and unique to each system. 

First, we contrast the qLR methods against each other before addressing the system- and state-dependent 
behaviour. 
For the full space systems, H$_2$ (SI \autoref{SI:fig:H2_save_or_not}) and H$_4$ (\autoref{fig:H4_save_or_not}), proj qLR has a larger $\sigma_k$ than naive qLR, while for active space simulations, LiH(2,2) (\autoref{fig:LiH_all}) and BeH$_2$(4,4) (SI \autoref{SI:fig:BeH2_save_or_not}), the difference between the methods is much smaller. We can see this replicated in some of the metrics introduced in \autoref{sec:theo:metrics} and shown in \autoref{SI:tab:metrics_H2}, \ref{tab:metrics_LiH}, \ref{tab:metrics_H4} and \ref{SI:tab:metrics_BeH2} for H$_2$, LiH(2,2), H$_4$ and BeH$_2$(4,4), respectively.
For H$_4$, $\overline{\sigma}_M$ and CV reveal that proj qLR has much larger (up to a factor of 10) matrix stds, confirming the sampled std results from the shot noise simulator. For LiH(2,2), the metrics are very similar between the qLR methods, again aligning with the sampled results. 
\begin{table}[htbp]
\caption{Quantum metrics for qLR of LiH(2,2) as described in \autoref{sec:theo:metrics}. The matrix $\boldsymbol{B}$ becomes (nearly) zero for proj and allproj qLR (cf. working equation Ref.\cite{Ziems2024}) leading to large and infinite cond and CV, respectively.}
\begin{center}
\begin{tabular}{l|rrrr}
 & \multicolumn{1}{c|}{\textbf{cond}} & \multicolumn{1}{c|}{$\overline{\sigma}_M$} & \multicolumn{1}{c|}{$\overline{\sigma}_{M,nc}$} & \multicolumn{1}{c}{\textbf{CV}} \\ \hline
 & \multicolumn{ 4}{c}{\textbf{naive qLR}} \\ \hline
$\boldsymbol{A}$ & 200 & 0.27 & 1.60 & 32.33  \\ 
$\boldsymbol{B}$ & 356 & 0.01 & 1.26 & 12.36 \\ 
$\boldsymbol{\Sigma}$ & 41 & 0.03 & 0.15 & 5.55  \\ 
$\textbf{E}^{[2]}$ & 232 &  &  &  \\ 
$(\textbf{S}^{[2]})^{-1}\textbf{E}^{[2]}$ & 49 &  &  &  \\ \hline
 & \multicolumn{ 4}{c}{\textbf{proj qLR}} \\ \hline
$\boldsymbol{A}$ & 201 & 0.36 & 2.34 & 36.06 \\
$\boldsymbol{B}$ & large & 0.06 & 1.85 & large \\
$\boldsymbol{\Sigma}$ & 41 & 0.03 & 0.22 & 5.42 \\
$\textbf{E}^{[2]}$ & 232 &  &  &  \\ 
$(\textbf{S}^{[2]})^{-1}\textbf{E}^{[2]}$ & 49 &  &  &  \\ \hline
 & \multicolumn{ 4}{c}{\textbf{allproj qLR}} \\ \hline
$\boldsymbol{A}$ & 202 & 0.37 & 3.00 & 36.69 \\
$\boldsymbol{B}$ & inf & 0.05 & 0.86 & inf \\ 
$\boldsymbol{\Sigma}$ & 41 & 0.03 & 0.22 & 5.42 \\
$\textbf{E}^{[2]}$ & 202 &  &  &  \\ 
$(\textbf{S}^{[2]})^{-1}\textbf{E}^{[2]}$ & 47 &  &  &  \\ 
\hline
\end{tabular}
\end{center}
\label{tab:metrics_LiH}
\end{table}
\begin{table}[htbp]
\caption{Quantum metrics for qLR of H$_4$ as described in \autoref{sec:theo:metrics}. The matrix $\boldsymbol{B}$ becomes zero for proj qLR (cf. working equation Ref.\cite{Ziems2024}) leading toinfinite cond and CV, respectively.}
\begin{center}
\begin{tabular}{l|rrrr}
 & \multicolumn{1}{c|}{\textbf{cond}} & \multicolumn{1}{c|}{$\overline{\sigma}_M$} & \multicolumn{1}{c|}{$\overline{\sigma}_{M,nc}$} & \multicolumn{1}{c}{\textbf{CV}} \\ \hline
 & \multicolumn{ 4}{c}{\textbf{naive qLR}} \\ \hline
$\boldsymbol{A}$ & 4.37 & 0.15 & 20.78 & 2.78 \\
$\boldsymbol{B}$ & 12.27 & 0.04 & 14.90 & 2.83 \\
$\boldsymbol{\Sigma}$ & 1.07 & 0.15 & 6.25 & 257.27 \\
$\textbf{E}^{[2]}$ & 6.71 &  &  &  \\ 
$(\textbf{S}^{[2]})^{-1}\textbf{E}^{[2]}$ & 6.59 &  &  &  \\ 
 & \multicolumn{ 4}{c}{\textbf{proj qLR}} \\ \hline
$\boldsymbol{A}$ & 4.52 & 0.75 & 258.49 & 25.33 \\
$\boldsymbol{B}$ & inf & 0.32 & 232.18 & inf \\
$\boldsymbol{\Sigma}$ & 1.08 & 0.14 & 25.11 & 436.17 \\
$\textbf{E}^{[2]}$ & 4.52 &  &  &  \\ 
$(\textbf{S}^{[2]})^{-1}\textbf{E}^{[2]}$ & 4.55 &  &  &  \\ 
\end{tabular}
\end{center}
\label{tab:metrics_H4}
\end{table}

Interestingly, for all systems the matrix std, 
$\overline{\sigma}_M$, in $\boldsymbol{A}$ dominates. As evident from \autoref{eq:qstd_2}, a matrix std's value comes from both the Pauli string variance and the coefficient. Comparing $\overline{\sigma}_M$ with $\overline{\sigma}_{M,nc}$ helps to differentiate this as the latter ignores the coefficient. From this, we can see that the dominance of $\boldsymbol{A}$ over $\boldsymbol{B}$ stems from the coefficient, while the dominance of $\boldsymbol{A}$ over $\boldsymbol{\Sigma}$ (if present) stems from the Pauli string variance. 
For example, for H$_4$ naive (proj) qLR in \autoref{tab:metrics_H4}, $\overline{\sigma}_M$ differs between $\boldsymbol{A}$ and $\boldsymbol{B}$ by a factor of 4 (2), while the coefficient free std, $\overline{\sigma}_{M,nc}$, are similar. On the other hand, $\boldsymbol{A}$ and $\boldsymbol{\Sigma}$ are identical (factor 2) in $\overline{\sigma}_M$, but differ by a factor of 3 (10) in $\overline{\sigma}_{M,nc}$.

Second, comparing the sampled std $\sigma_k$ between the four systems, one observes that \textit{(i)} the full space systems have smaller std than their active space counterparts and \textit{(ii)} that the std increases with qubit size. Thus, we observe that the sampled std increases from H$_2$, to H$_4$, LiH(2,2), and BeH$_2$(4,4). 
While the condition number, cond, does not seem to be a reliable indicator for the subtle differences between qLR methods for a given system, it correctly predict this trend between systems. For example, for naive qLR the condition number of the Hessian, $\textbf{E}^{[2]}$, rises from $2.15,6.71,232,960$ for H$_2$, H$_4$, LiH(2,2), BeH$_2$(4,4), respectively. The inter-system trend is also captured by the CV of the dominant $\boldsymbol{A}$ matrix.

Third, we come back to the state-dependent behaviour of the sampled std, $\sigma_k$, and focus on LiH(2,2) (see SI for others). In \autoref{fig:LiH_all} (top row), we see the sampled std for each state. The most interesting feature is the dramatic decrease in std for excited states $k=8,9,10$.
In \autoref{fig:LiH_all} (bottom row), we see the state-specific std, $\overline{\sigma}_{M,k}$, for LiH that reconstructs the noise contribution to each state stemming from each qLR matrix (cf. \autoref{sec:theo:metrics}). It reproduces the overall trends of the actual sampled std, 
$\sigma_k$, in the case of no PS (solid line, top row \autoref{fig:LiH_all}), importantly, including the dip for the excited states $k=8,9,10$. This dip can now be explained as these excitation energies are dominated by transitions that originate from inactive-to-virtual orbital rotations. 
Our state-specific std approach can capture all this (see \autoref{SI:sec:add_shotnoise} for other systems) since it is based on classic quantum chemistry transition analysis of excited states, combined with the std of individual quantum operators based on their Pauli string decomposition. 

Not all subtle differences within one order of magnitude of std are captured, but large trends are perfectly reproduced (see also SI \autoref{SI:sec:add_shotnoise}). To understand the discrepancies, we recall that the matrix and state-specific std \textit{(i)} only account for trends in a given qLR matrix and not in the final excitation energies, and \textit{(i)} assigns approximately a std to each quantum operator by row-wise integration of the qLR matrices.

As discussed in \autoref{sec:results:PS}, Pauli Saving leads to a reduction in measurement and noise, but \autoref{fig:LiH_all} (top row) shows that, additionally, the trend between excited states is changed. This change cannot be captured by $\overline{\sigma}_{M,k}$ (bottom row). However, the overall std behaviour between the different qLR methods and different systems is preserved with PS.

In summary, this shows that the metrics that are directly based on the std of Pauli strings can explain how the quantum mechanical shot noise of quantum computers impact each level of the qLR algorithms differently and is parametrization-, system-, and excited state-dependent. It reveals that simple methods used so far like adding Gaussian distributed random numbers as noise to matrix elements\cite{Asthana2023,Kumar2023} cannot capture these subtleties and thus will not give results applicable to real quantum computers. We also note that the chosen 100k shots per Pauli string on simulated fault-tolerant quantum computers is not sufficient to reach chemical accuracy in the standard deviation for all excited states and systems presented. Again, it shows that even for small systems high measurement numbers (and speed) will be crucial and limiting even in the fault-tolerant regime.


\subsection{Hardware results}
Having understood the impact of shot noise on the qLR results, we move to real hardware simulations on IBM Osaka for LiH(2,2). On near-term devices like this, we have additional device noise that leads to systematic and random noise. The latter is expected to broaden the standard deviation known from the shot noise study above, while systematic noise leads to a bias/shift away from the real mean of the expectation values. To mitigate these, we use our Ansatz-based read-out and gate error mitigation introduced in \autoref{sec:theo:MA0}.
\begin{figure}[htbp]
    \centering 
    \includegraphics[width=\textwidth]{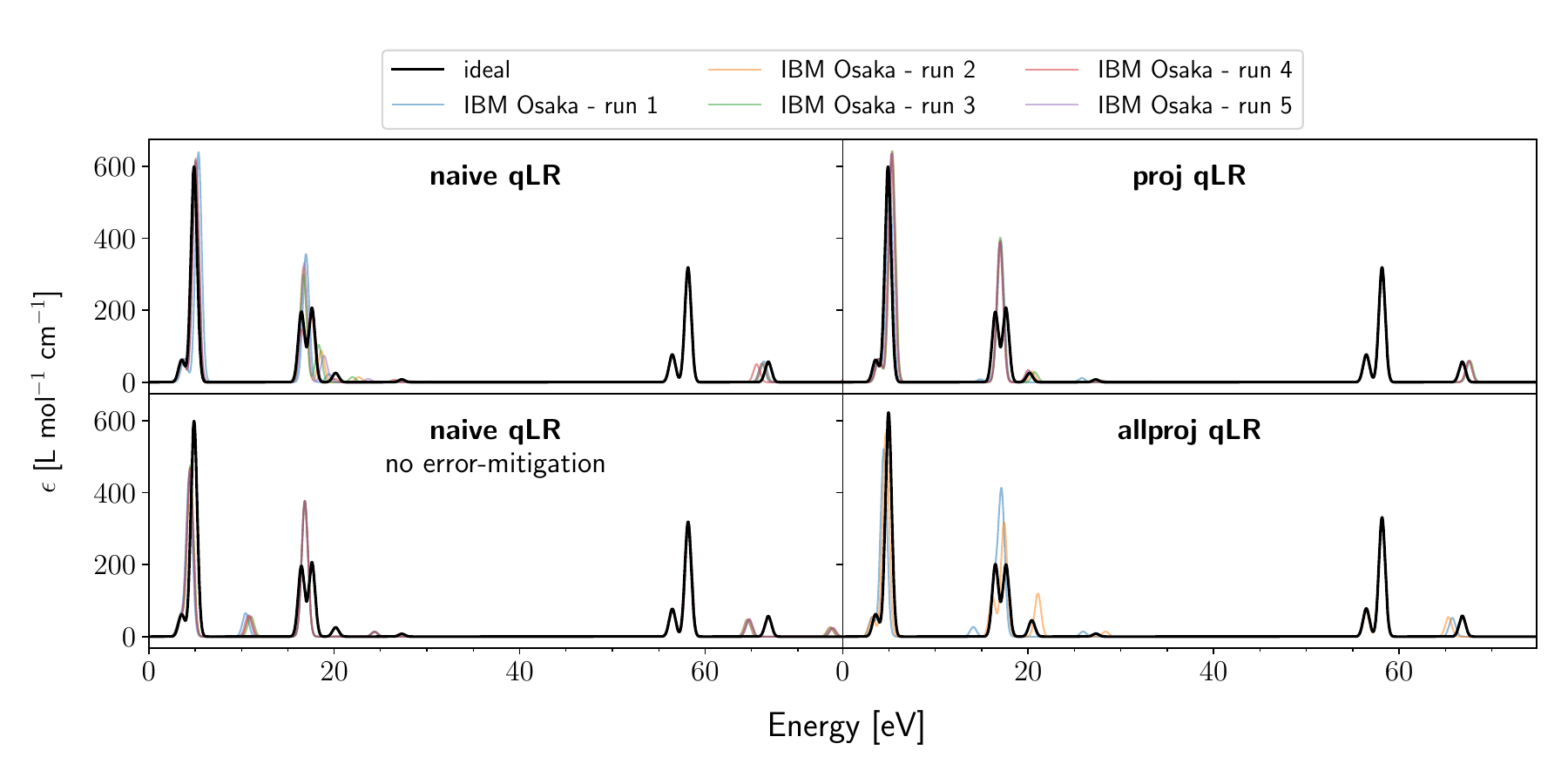}
    \caption{Absorption spectra of LiH(2,2) / STO-3G calculated on IBM Osaka with various qLR methods. The black line indicates the ideal (shot- and device-noise free) results that coincide with classic CASSCF. For each qLR method, five quantum hardware runs were performed and are shown in different colours. For allproj qLR, quantum run 1, 3, and 5 failed as they produced negative Hessian eigenvalues and were removed at post-processing. Detailed and zoomed-in results of each hardware run are found in SI \autoref{SI:sec:add_hardware}.}
    \label{fig:LiH_hardware}
\end{figure}

In \autoref{fig:LiH_hardware}, we present hardware results for all three qLR methods in the STO-3G basis set. The black line represents the ideal result (shot- and device-noise free) that is identical to classic CASSCF LR\cite{Ziems2024}. In order to judge reproducibility, the hardware simulations were run 5 times. It shows that not all peaks are captured equally well. The peaks in the area of around $15-20\,$eV show the largest deviations, while the peaks around $55-60\,$eV are perfectly matched every time. This fully agrees with the shot noise study above (see \autoref{fig:LiH_all} and SI \autoref{SI:fig:LiH_save_or_not}), where these two areas have, respectively, the largest and smallest shot noise-induced deviations. This confirms that on near-term hardware the additional device noise leads to a broadening of the quantum mechanical shot noise. Between all three methods, none performs clearly better as expected from our quantum metrics analysis above. However, allproj qLR consists only of two runs as the three other runs were removed in post-processing as they had negative Hessian eigenvalues. Based on the analysis performed in the previous section and the low statistics count of only 5 hardware runs per qLR method, we do not believe that this is characteristic of allproj qLR but rather shows that the error rate of quantum computers can suddenly change between runs leading to nonphysical results even with error mitigation present. 

Furthermore, \autoref{fig:LiH_hardware} also shows the effect of our Ansatz-based gate and read-out error mitigation. The first column reveals the effect for naive qLR. Systematic hardware errors that lead to shifts in peaks are clearly mitigated by this approach. 

For each qLR run, the error mitigation had a cost of 2 million shots with a quantum runtime of $9\,$min, while the qLR simulation itself had a cost of 4.5 million shots taking $20\,$min each. This shows how costly and time-consuming even simple systems are on current near-term devices. 

\begin{figure}[ht]
    \centering 
    \includegraphics[width=\textwidth]{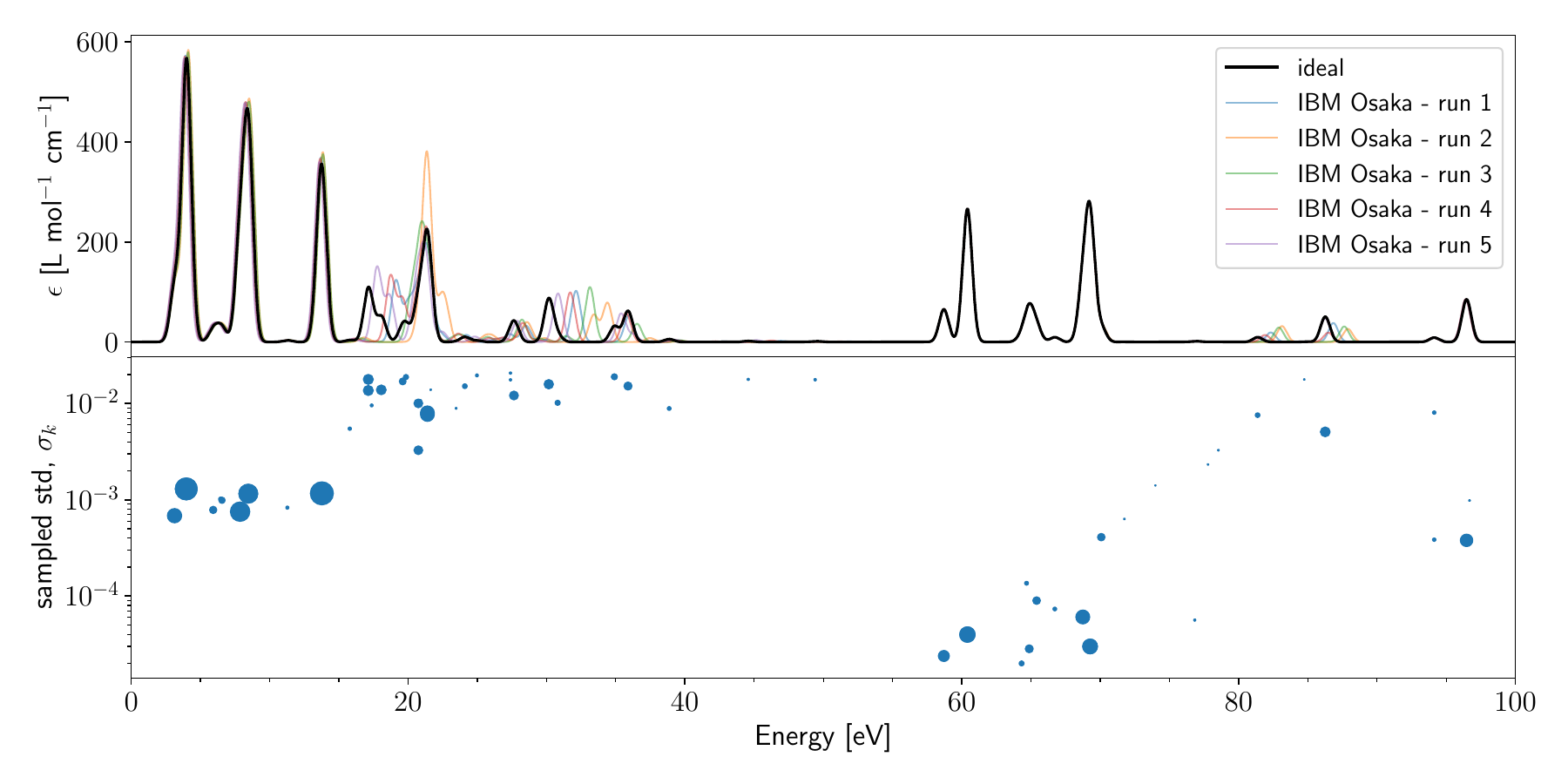}
    \caption{Top: Absorption spectra of LiH(2,2) / cc-pVTZ calculated on IBM Osaka with naive qLR and 1M shots per Pauli string. The black line indicates the ideal (shot- and device-noise free) results that coincide with classic CASSCF. Five quantum hardware runs were performed and are shown in different colours. Detailed results of each hardware run are found in \autoref{SI:sec:add_hardware}. Bottom: Sampled standard deviation, $\sigma_k$, for 1000 samples with 100k shots on shot noise simulator. The size of the dot correlates with the oscillator strength value.}
    \label{fig:LiH_hardware_TZ}
\end{figure}
Lastly, in \autoref{fig:LiH_hardware_TZ} (and in SI \autoref{SI:fig:LiH_hardware_TZ_500k}), we present naive qLR LiH(2,2) with cc-pVTZ basis set using 1M (500k) shots per Pauli string. The bottom panel show the sampled std, $\sigma_k$, for LiH(2,2) / cc-pVTZ on the shot noise simulator and correlates the dot size with the oscillator strength. As for the STO-3G case, there is a clear correlation between std from shot noise and deviation of the hardware runs from ideal results. In general, the STO-3G results show less deviation than the cc-pVTZ results (compare also \autoref{SI:fig:LiH_save_or_not} and \ref{SI:fig:LiH_naive_TZ_std} in the SI). We can reason this with our quantum metrics (see SI \autoref{SI:tab:metrics_LiH_TZ}) as the triple-zeta simulation has larger cond (1199 for $\textbf{E}^{[2]}$, 111 for $(\textbf{S}^{[2]})^{-1}\textbf{E}^{[2]}$) and CV (51 for $\boldsymbol{A}$, 33 for $\boldsymbol{B}$, 27 for $\boldsymbol{\Sigma}$) than the single-zeta metrics in \autoref{tab:metrics_LiH}. This is due to the cc-pVTZ basis set being more diffuse than the STO-3G basis set.
This causes some of the MO-coefficient to be small, which increases the sensitivity to noise.

\section{\label{sec:summary}Summary}

In this work, we presented a first-of-its-kind detailed study of quantum linear response theory on simulated fault-tolerant QPUs and current near-term quantum hardware. Specifically, we presented quantum metrics to understand and analyze the origin of noise in our algorithms, introduced Ansatz-based error mitigation, and revealed the impact of Pauli saving. Importantly, the findings reported are applicable to many hybrid quantum-classical algorithms in chemistry and provide guidance for choosing, understanding, and executing similar works in the future. 

To this end, we studied H$_2$, H$_4$, LiH and BeH$_2$ on a shot noise simulator, i.e.\ a simulated fault-tolerant quantum computer, and discussed how the quantum mechanical shot noise of these devices impacts differently depending on the algorithm, system, and even the specific excited state. Using a standard deviation metric based on the individual Pauli string distribution, we reveal the coefficient of variation as a reliable indicator to compare algorithms across and within systems. Moreover, the unique standard deviation of each operator translates to different deviations in the individual excitation peak. Interestingly, we show that Pauli saving leads to a reduction in both measurement cost and noise by up to two orders of magnitude. This effect is likely to increase for larger systems.

The hardware results of LiH using up to cc-pVTZ basis set are a first small example of quantum simulations of spectroscopic properties towards usable results beyond the minimal basis approach commonly adopted in quantum computing for chemistry. The qLR algorithm in combination with the chosen tUCCSD ansatz can replicate classic quantum chemistry (CASSCF) for LiH. The results show that our Ansatz-based error mitigation and a large number of shots can capture most features of the absorption spectrum but accuracy still limited by measurement speed and error rates. We confirm that the excitation energy trend found in shot noise simulation and studied with our error metrics translates to quantum hardware.

This work is understood as a proof-of-principle to show that molecular excited state properties can be obtained on a quantum computer in a systematic and general fashion that is inspired by classic quantum chemistry and does not rely on approximated or hand-crafted circuits, pre-knowledge of the classic solution, massive classic post-processing or full state tomography allowing for density matrix clean-up. 
We show that for very small systems near-term hardware results can converge to classic state-of-the-art quantum chemistry accuracy. In that goal, this work diverges from many hardware studies that focus on large qubit number simulations in minimal basis sets or approximated Ans\"atze, where the results often fall short of the accuracy achieved by classical quantum chemistry, sometimes merely replicating mean-field solutions.
Importantly, this work also showcases that, in order for quantum computers to have any application in chemistry and scale simulations with accurate, multi-configurational results using proper basis sets to meaningful system sizes, let alone beyond classic capabilities, great improvements in hardware error rates and measurement speed are needed. 
Nonetheless, while quantum computing for chemistry is in its infancy, the prospect of circumventing the exponential scaling of quantum chemistry algorithms is highly 
impactful and motivates detailed fundamental research efforts. 

\newpage
\section*{Acknowledgments}
Financial Support from the Novo Nordisk Foundation (NNF) for the focused research project ``Hybrid Quantum Chemistry on Hybrid Quantum Computers'' (NNF grant NNFSA220080996) is acknowledged. 


\bibliography{literature}

\end{document}


\newpage

\section{\label{sec:add_shotnoise}Shot noise simulations: Additional Tables and Figures}

\begin{table}[htbp]
\caption{For different qLR implementations, the number of Pauli strings to be evaluated using Pauli saving (PS) and qubit-wise commutivity (QWC) for different molecules in minimal basis. Proj-ff and allproj-ff refer to a full fermionic (ff) implementation of proj and allproj qLR, respectively, where all fermionic terms in the working equation are calculated even if they appear multiple times (see appendix in Ref.~\citenum{main:Ziems2024}).}
\begin{center}
\begin{tabular}{l|ccc}
\multicolumn{1}{c|}{\textbf{}} & \multicolumn{1}{c|}{\textbf{PS + QWC}} & \multicolumn{1}{c|}{\textbf{QWC}} & \multicolumn{1}{c}{\textbf{none}} \\ \hline
 & \multicolumn{ 3}{c}{\textbf{H}$\boldsymbol{_2}$} \\ \hline
\textbf{naive} & 9 & 35 & 42 \\ 
\textbf{proj} & 9 & 38 & 64 \\ 
\textbf{proj-ff} & 9 & 171 & 275 \\ \hline
 & \multicolumn{ 3}{c}{\textbf{LiH(2,2)}} \\ \hline
\textbf{naive} & 9 & 1118 & 1774 \\ 
\textbf{proj} & 9 & 922 & 1491 \\ 
\textbf{allproj} & 9 & 447 & 715 \\ 
\textbf{proj-ff} & 9 & 1085 & 1730 \\ 
\textbf{allproj-ff} & 9 & 614 & 963 \\ \hline
 & \multicolumn{ 3}{c}{\textbf{H}$\boldsymbol{_4}$} \\ \hline
\textbf{naive} & 699 & 22908 & 76406 \\ 
\textbf{proj} & 774 & 15941 & 83041 \\ 
\textbf{proj-ff} & 752 & 72990 & 280893 \\ \hline
 & \multicolumn{ 3}{c}{\textbf{BeH}$\boldsymbol{_2}$\textbf{(4,4)}} \\ \hline
\textbf{naive} & 822 & 104096 & 420132 \\ 
\textbf{proj} & 753 & 64076 & 309531 \\ 
\textbf{allproj} & 753 & 44510 & 227781 \\ 
\textbf{proj-ff} & 822 & 147275 & 662395 \\ 
\textbf{allproj-ff} & 790 & 127745 & 580810 \\ \hline
\end{tabular}
\end{center}
\label{tab:PS}
\end{table}

\newpage

\subsection{H\texorpdfstring{$_2$}{2}}

\begin{figure}[htbp!]
    \centering 
    \includegraphics[width=.7\textwidth]{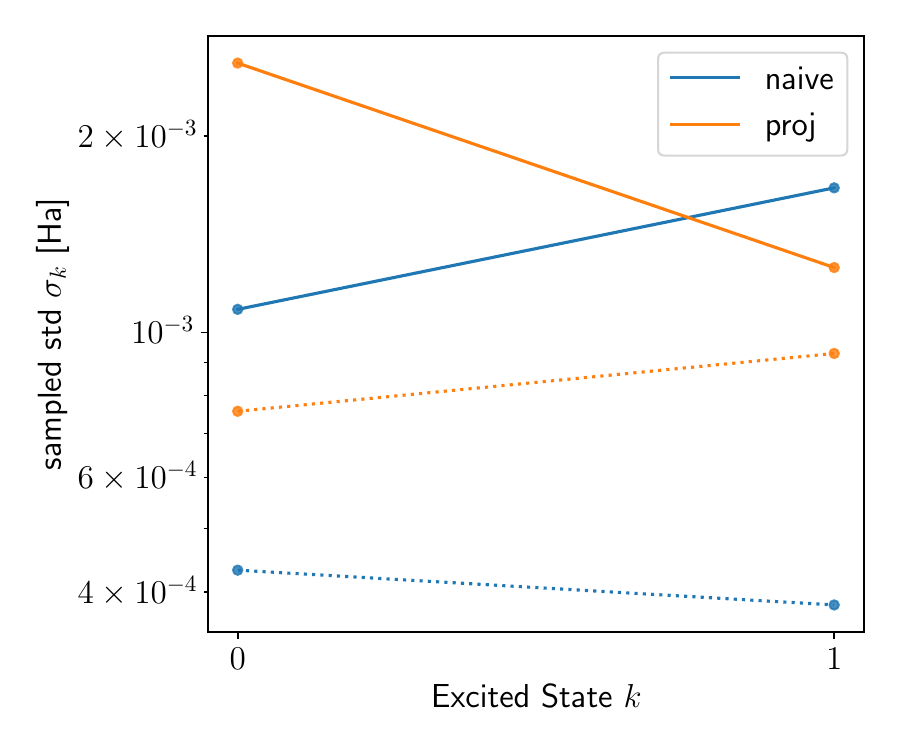}
    \caption{Sampled std, $\sigma_k$, with (without) Pauli saving (PS) in dotted (solid) line for H$_2$ naive and proj qLR.}
    \label{fig:H2_save_or_not}
\end{figure}

\begin{figure}[htbp!]
    \centering 
    \includegraphics[width=.7\textwidth]{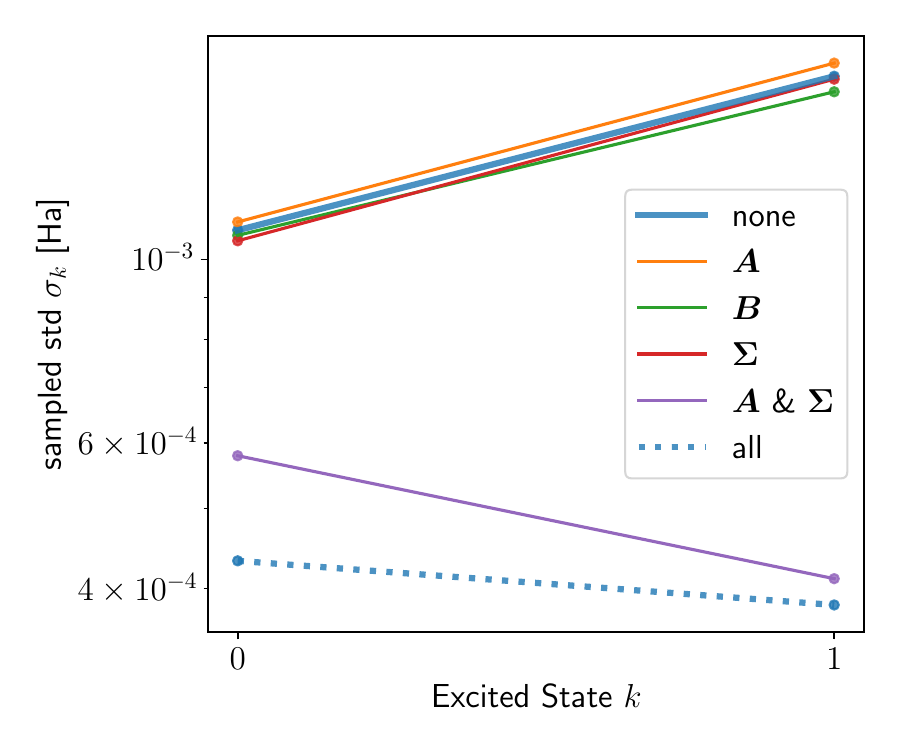}
    \caption{Sampled std, $\sigma_k$, for H$_2$ naive qLR with using Pauli saving (PS) for only selected matrices in the qLR equations.}
    \label{fig:H2_save_selected}
\end{figure}

\begin{figure}[htbp!]
    \centering 
    \includegraphics[width=.7\textwidth]{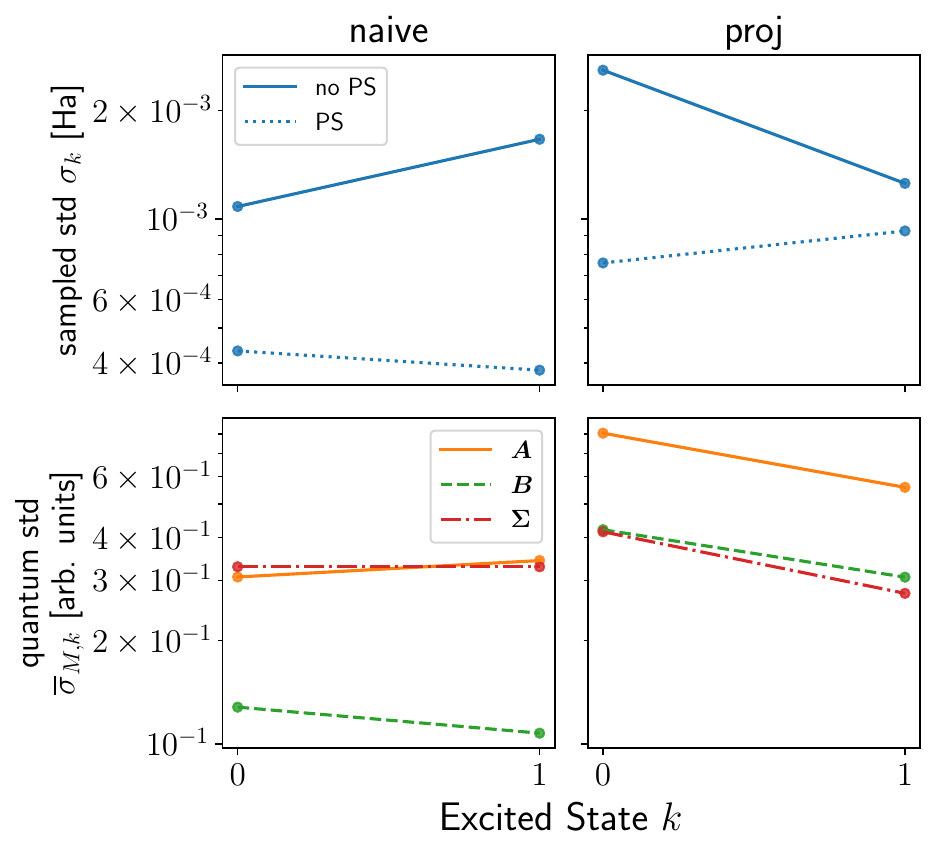}
    \caption{Standard deviation (std) analysis of naive (left), and proj (right) qLR of H$_2$. The top row shows the samples std, $\sigma_k$, with (dotted line) and without (solid line) Pauli saving (PS). The bottom row shows the state-speciifc std, $\overline{\sigma}_{M,k}$ for the qLR matrices $\boldsymbol{M}=\boldsymbol{A}$ (orange solid), $\boldsymbol{M}=\boldsymbol{B}$ (green dashed) and $\boldsymbol{M}=\boldsymbol{\Sigma}$ (red dashed-dotted).}
    \label{fig:H2_all}
\end{figure}

For H$_2$ proj qLR observes larger sampled std  $\sigma_k$  than naive qLR (see \autoref{fig:H2_save_or_not} and \ref{fig:H2_all} (top row)). This is replicated in the state-specific std, $\overline{\sigma}_{M,k}$, in \autoref{fig:H2_all} (bottom row), and the matrix std, $\overline{\sigma}_M$, and CV in \autoref{tab:metrics_H2}. The condition number, cond, is no reliable metric for this. In SI \autoref{fig:H2_save_selected}, we see that the effect of PS is mainly driven by a combined PS in $\boldsymbol{A}$ and $\boldsymbol{\Sigma}$. This shows the importance of PS for the inversion step of a generalized eigenvalue problem, see \autoref{sec:PS}.

\begin{table}[htbp]
\caption{Quantum metrics for qLR of H$_2$.}
\begin{center}
\begin{tabular}{l|rrrr}
 & \multicolumn{1}{c|}{\textbf{cond}} & \multicolumn{1}{c|}{$\overline{\sigma}_M$} & \multicolumn{1}{c|}{$\overline{\sigma}_{M,nc}$} & \multicolumn{1}{c}{\textbf{CV}} \\ \hline
 & \multicolumn{ 4}{c}{\textbf{naive qLR}} \\ \hline
$\boldsymbol{A}$ & 1.63 & {0.32} & {2.41} & {0.19} \\ 
$\boldsymbol{B}$ & 5.39 & {0.12} & {1.99} & {1.94} \\ 
$\boldsymbol{\Sigma}$ & 1.00 & {0.33} & {1.16} & {0.16} \\ 
$\textbf{E}^{[2]}$ & 2.15 &  &  &  \\ 
$(\textbf{S}^{[2]})^{-1}\textbf{E}^{[2]}$ & 2.15 &  &  &  \\ \hline
 & \multicolumn{ 4}{c}{\textbf{proj qLR}} \\ \hline
$\boldsymbol{A}$ & 1.65 & {0.68} & {11.05} & {0.59} \\ 
$\boldsymbol{B}$ & inf & {0.36} & {9.50} & {inf} \\ 
$\boldsymbol{\Sigma}$ & 1.01 & {0.34} & {4.47} & {0.30} \\ 
$\textbf{E}^{[2]}$ & 1.65 &  &  &  \\ 
$(\textbf{S}^{[2]})^{-1}\textbf{E}^{[2]}$ & 1.67 &  &  &  \\\hline
\end{tabular}
\end{center}
\label{tab:metrics_H2}
\end{table}

\newpage

\subsection{LiH}

\begin{figure}[htbp]
    \centering 
    \includegraphics[width=.7\textwidth]{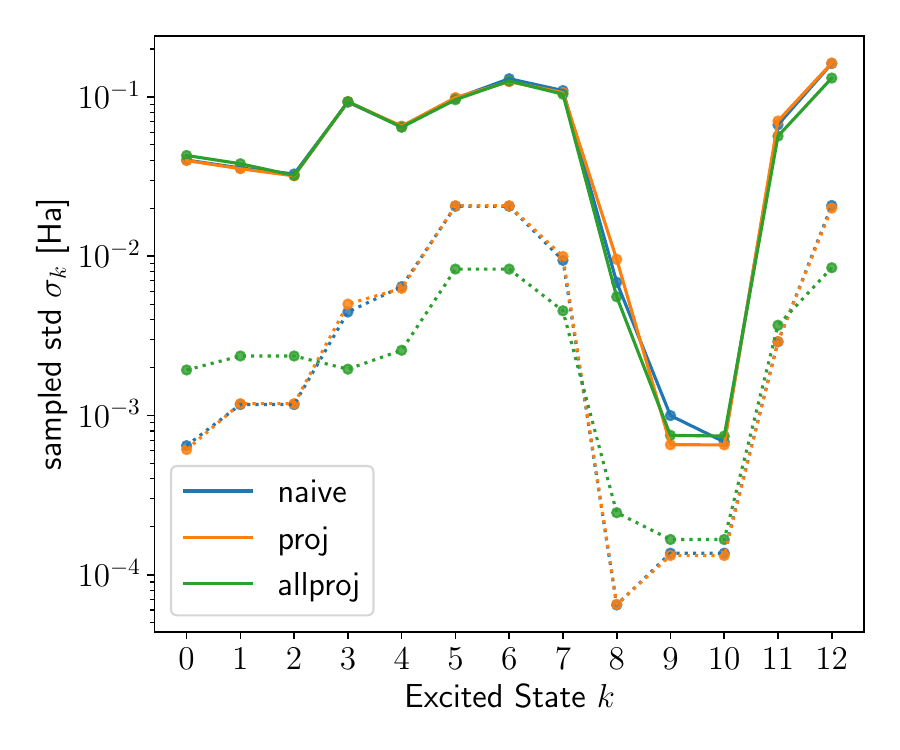}
    \caption{Sampled std, $\sigma_k$, with (without) Pauli saving (PS) in dotted (solid) line for LiH(2,2) naive, proj and allproj qLR.}
    \label{fig:LiH_save_or_not}
\end{figure}

\begin{figure}[htbp]
    \centering 
    \includegraphics[width=.7\textwidth]{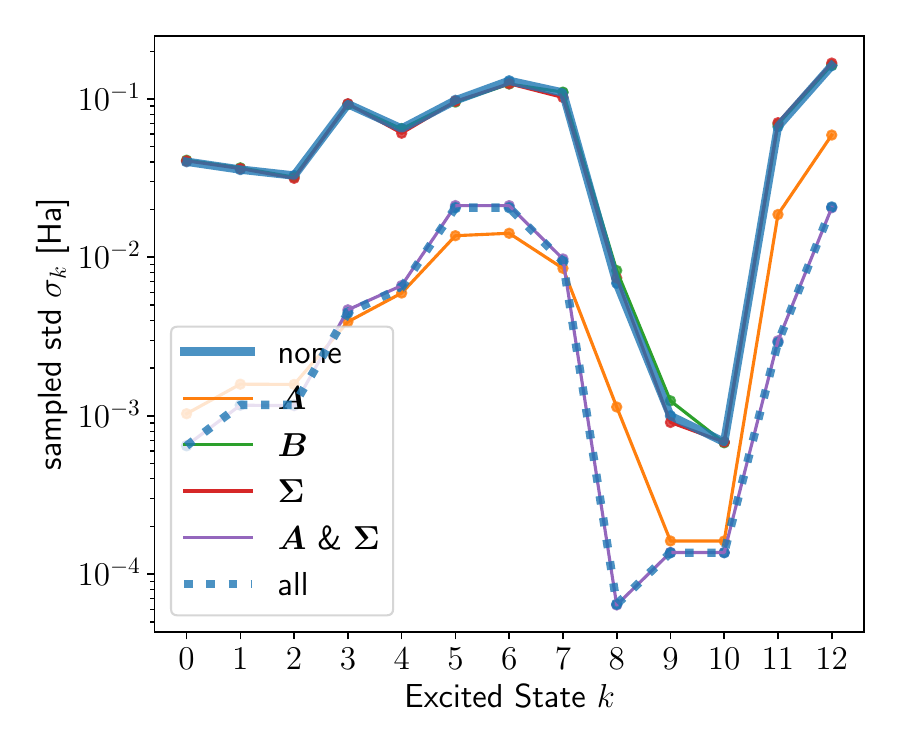}
    \caption{Sampled std $\sigma_k$ for LiH(2,2) naive qLR combined with using Pauli saving (PS) for only selected matrices in the qLR equations.}
    \label{fig:LiH_save_selected}
\end{figure}

In addition to the figures and discussion in the text, we present a combined figure of the sampled std, $\sigma_k$, in SI \autoref{fig:LiH_save_or_not} to visualise the effect of PS and how equal the std between the parametrizations is without PS. Additionally, SI \autoref{fig:LiH_save_selected} reveals that both PS only on $\boldsymbol{A}$ and PS on $\boldsymbol{A}$ and $\boldsymbol{\Sigma}$ drive the effect of noise-reduction via PS. This shows that both inversion step and eigenvalue diagonalization (as presented in SI \autoref{sec:PS}) are important to explain PS. 

\begin{figure}[htbp]
    \centering 
    \includegraphics[width=.7\textwidth]{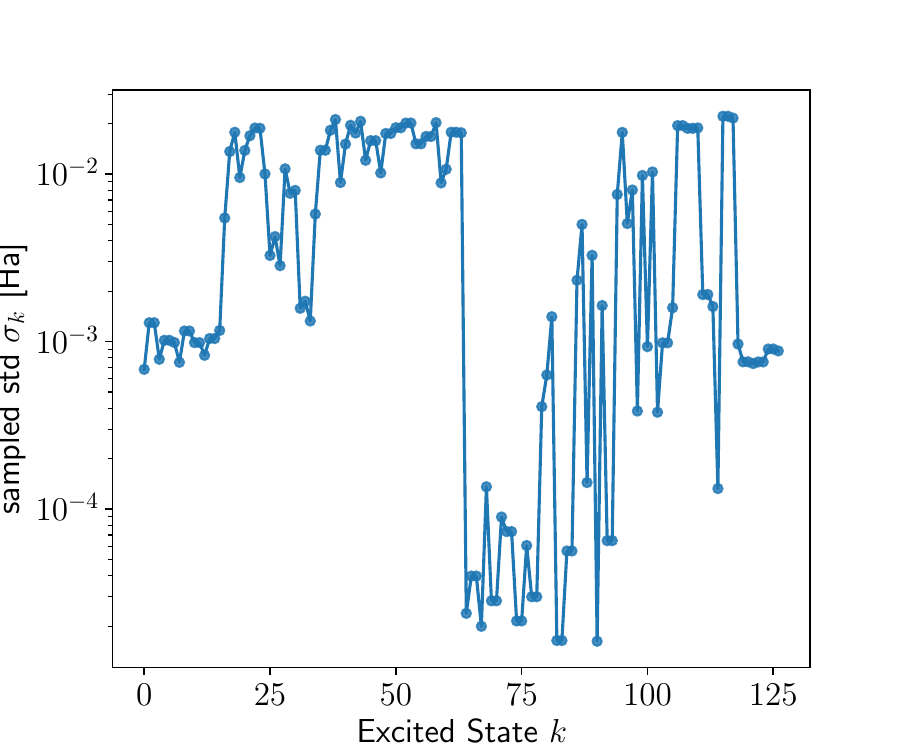}
    \caption{Sampled std $\sigma_k$ with Pauli saving for LiH(2,2) naive qLR with cc-pVTZ.}
    \label{fig:LiH_naive_TZ_std}
\end{figure}

In \autoref{fig:LiH_naive_TZ_std}, we show the sampled std, $\sigma_k$, for cc-pVTZ per excited state as alternative visualization to the main text \autoref{main:fig:LiH_hardware_TZ} (bottom row). \autoref{tab:metrics_LiH_TZ} presents the metrics for the triple-zeta LiH naive qLR. Compared to the STO-3G version in \autoref{main:tab:metrics_LiH} (main text) the metrics suitable for inter-system comparison (cond and CV) are much larger explaining the increased sampled std.

\begin{table}[htbp]
\caption{Quantum metrics for qLR of LiH(2,2) / cc-pVTZ as described in \autoref{main:sec:theo:metrics}.}
\begin{center}
\begin{tabular}{l|rrr}
 & \multicolumn{1}{c|}{\textbf{cond}} & \multicolumn{1}{c|}{$\overline{\sigma}_M$} &  \multicolumn{1}{c}{\textbf{CV}} \\ \hline
 & \multicolumn{ 3}{c}{\textbf{naive qLR}} \\ \hline
$\boldsymbol{A}$ & 652 & 0.029 & 50.85  \\ 
$\boldsymbol{B}$ & large & 0.001  & 33.40 \\ 
$\boldsymbol{\Sigma}$ & 46 & 0.002 & 26.71  \\ 
$\textbf{E}^{[2]}$ & 1199 &   &  \\ 
$(\textbf{S}^{[2]})^{-1}\textbf{E}^{[2]}$ & 111 &  &  \\ \hline
\end{tabular}
\end{center}
\label{tab:metrics_LiH_TZ}
\end{table}

\newpage

\subsection{H\texorpdfstring{$_4$}{4}}

\begin{figure}[h!]
    \centering 
    \includegraphics[width=0.7\textwidth]{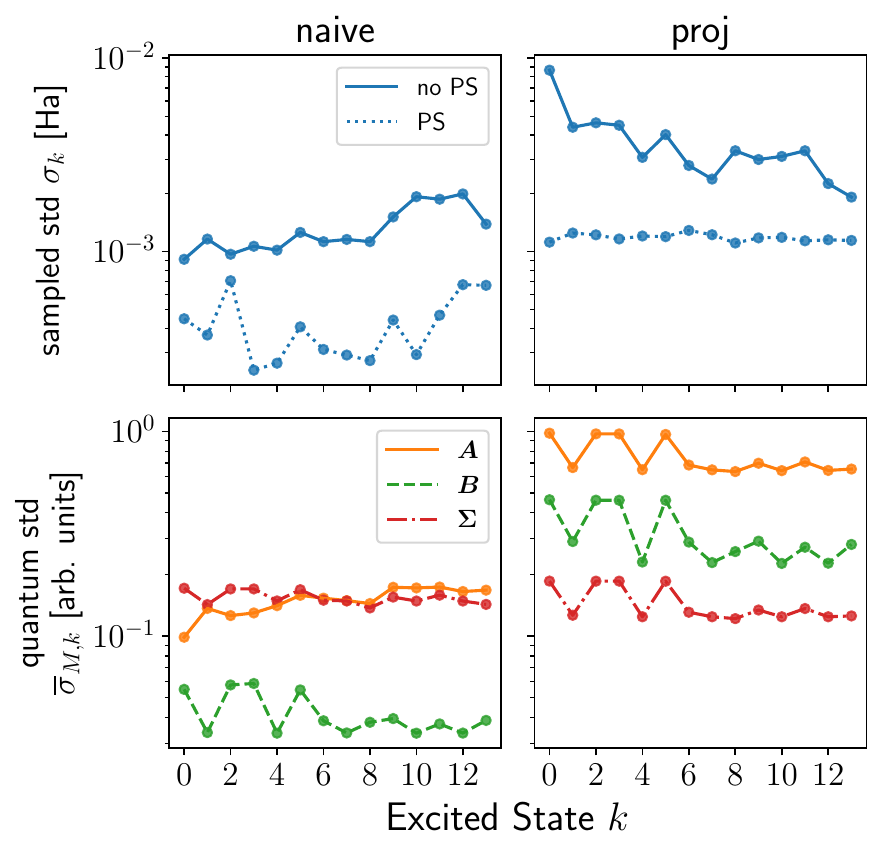}
    \caption{Standard deviation (std) analysis of naive (left), and proj (right) qLR of H$_4$. The top row shows the samples std, $\sigma_k$, with (dotted line) and without (solid line) Pauli saving (PS). The bottom row shows the state-speciifc std, $\overline{\sigma}_{M,k}$ for the qLR matrices $\boldsymbol{M}=\boldsymbol{A}$ (orange solid), $\boldsymbol{M}=\boldsymbol{B}$ (green dashed) and $\boldsymbol{M}=\boldsymbol{\Sigma}$ (red dashed-dotted).}
    \label{fig:H4_all}
\end{figure}

In addition to the results for H$_4$ in the main text, SI \autoref{fig:H4_all} shows that $\overline{\sigma}_{M,k}$ can again predict the major trends in the sampled std. 

\newpage

\subsection{BeH\texorpdfstring{$_2$}{2}}

\begin{figure}[htbp!]
    \centering 
    \includegraphics[width=.7\textwidth]{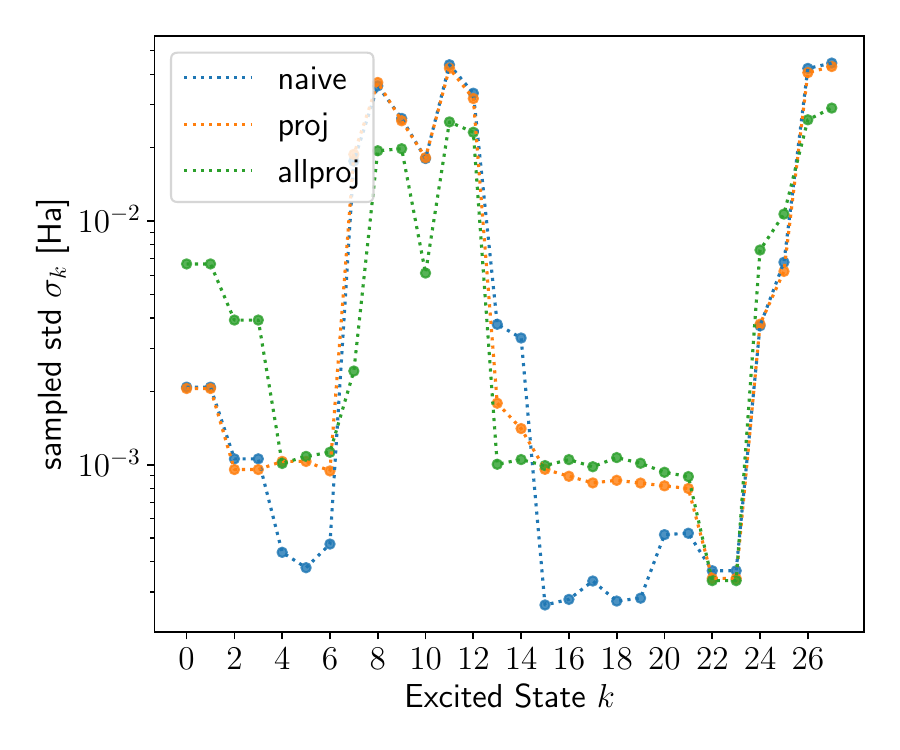}
    \caption{Sampled std, $\sigma_k$, with Pauli saving (PS) in dotted line for BeH$_2$(4,4) naive, proj and allproj qLR. Without PS, $>90\%$ of all runs lead to negative eigenvalues in the Hessian.}
    \label{fig:BeH2_save_or_not}
\end{figure}

As reported in the text, for BeH$_2$ $>90\,$\% of the runs without PS resulted in negative Hessian eigenvalues. Thus, in \autoref{fig:BeH2_save_or_not}, only results with PS are presented. 

\begin{table}[htbp]
\caption{Quantum metrics for qLR of BeH$_2$(4,4).}
\begin{center}
\begin{tabular}{l|rrrr}
 & \multicolumn{1}{c|}{\textbf{cond}} & \multicolumn{1}{c|}{$\overline{\sigma}_M$} & \multicolumn{1}{c|}{$\overline{\sigma}_{M,nc}$} & \multicolumn{1}{c}{\textbf{CV}} \\ \hline
 & \multicolumn{ 4}{c}{\textbf{naive qLR}} \\ \hline
$\boldsymbol{A}$ & 707 & 0.34 & 16.21 & 74 \\
$\boldsymbol{B}$ & 159 & 0.01 & 11.96 & 52 \\
$\boldsymbol{\Sigma}$ & 135 & 0.05 & 1.69 & 1047 \\
$\boldsymbol{E}^{[2]}$ & 960 &  &  &  \\ 
$(\boldsymbol{S}^{[2]})^{-1}\boldsymbol{E}^{[2]}$ & 73 &  &  &  \\ 
 & \multicolumn{ 4}{c}{\textbf{proj qLR}} \\ \hline
$\boldsymbol{A}$ & 707 & 1.18 & 154.72 & 1034 \\
$\boldsymbol{B}$ & large & 0.35 & 138.20 & large \\
$\boldsymbol{\Sigma}$ & 135 & 0.05 & 6.44 & 1740 \\
$\boldsymbol{E}^{[2]}$ & 960 &  &  &  \\ 
$(\boldsymbol{S}^{[2]})^{-1}\boldsymbol{E}^{[2]}$ & 73 &  &  &  \\ 
 & \multicolumn{ 4}{c}{\textbf{allproj qLR}} \\ \hline
$\boldsymbol{A}$ & 707 & 1.20 & 155.97 & 1035 \\
$\boldsymbol{B}$ & inf & 0.35 & 133.12 & inf \\
$\boldsymbol{\Sigma}$ & 135 & 0.05 & 6.44 & 1740 \\
$\boldsymbol{E}^{[2]}$ & 707 &  &  &  \\ 
$(\boldsymbol{S}^{[2]})^{-1}\boldsymbol{E}^{[2]}$ & 55 &  &  &  \\ 
\end{tabular}
\end{center}
\label{tab:metrics_BeH2}
\end{table}

\newpage

\section{\label{sec:PS}How Pauli saving leads to less noise}
In the following, we make a crude mathematical argument of why Pauli saving leads to less noise for the example of a 2x2 matrix.

\subsection{Part I: Simple matrix}

We have an eigenvalue equation:
\begin{align}
    Ax = \lambda  x
\end{align}
which has for a 2x2 matrix 
\begin{align}
    A &= \begin{pmatrix}
        a_{00} & a_{01} \\
        a_{10} & a_{11}
    \end{pmatrix} 
\end{align}
the eigenvalues:
\begin{align}
    \lambda_{1/2} = \frac{1}{2} (a_{00} + a_{11}) \pm \sqrt{(a_{00}-a_{11})^2+4a_{01}a_{10}}.
\end{align}

Next, we assume that each value consists of the ideal value and a deviation due to some kind of noise:
\begin{align}
    a_{ij} = \tilde a_{ij} + \sigma_{ij}.
\end{align}
This gives the eigenvalues
\begin{align}
    \lambda_{1/2} &= \frac{1}{2} \left[ ((a_{00}+\sigma_{00}) + (a_{11}+\sigma_{11})) \pm \sqrt{((a_{00}+\sigma_{00})-(a_{11}+\sigma_{11}))^2+4(a_{01}+\sigma_{01})(a_{10}+\sigma_{10})} \right] \nonumber \\
    &=\frac{1}{2} \left[ (\underbrace{a_{00} + a_{11}}_\text{value} + \underbrace{\sigma_{00}+\sigma_{11}}_\text{deviation}) \pm \sqrt{(\underbrace{a_{00}-a_{11}}_\text{value}+\underbrace{\sigma_{00}-\sigma_{11}}_\text{deviation})^2+4(\underbrace{a_{01}a_{10}}_\text{value}+ \underbrace{a_{01}\sigma_{10} + a_{10}\sigma_{01} +\sigma_{01}\sigma_{10}}_\text{deviation})}\right] \label{eq:simplemat_before}
\end{align}

We now invoke the limit case of equal noise, i.e.
\begin{align}
    \sigma_{00} = \sigma_{11} = \sigma_{01} = \sigma_{10}= \sigma,
\end{align}
which gives the following eigenvalues
\begin{align}
    \lambda_{1/2} &=\frac{1}{2} \left[ (\underbrace{a_{00} + a_{11}}_\text{value} + \underbrace{2\sigma}_\text{deviation}) \pm \sqrt{(\underbrace{a_{00}-a_{11}}_\text{value})^2+4(\underbrace{a_{01}a_{10}}_\text{value}+ \underbrace{a_{01}\sigma + a_{10}\sigma +\sigma^2}_\text{deviation})}\right]
\end{align}
Clearly, in this case the second deviation part from \autoref{eq:simplemat_before} vanishes.

\subsection{Part II: With inversion}

We have a general eigenvalue equation:
\begin{align}
    Ax = \lambda B x
\end{align}
that can be rearranged into:
\begin{align}
    B^{-1}Ax = \lambda x.
\end{align}

Let's look at this for 2x2 matrices defined as:
\begin{align}
    A &= \begin{pmatrix}
        a_{00} & a_{01} \\
        a_{10} & a_{11}
    \end{pmatrix} \\
    B &= \begin{pmatrix}
        b_{00} & b_{01} \\
        b_{10} & b_{11}
    \end{pmatrix}.
\end{align}
The inverse matrix, $B^{-1}$, is obtained as:
\begin{align}
    B^{-1} &= \frac{1}{\det B}\begin{pmatrix}
        b_{11} & - b_{01} \\
        - b_{10} & b_{00}
    \end{pmatrix} 
\end{align}
with 
\begin{align}
    \det B &=  b_{00} b_{11} - b_{01}b_{10}.
\end{align}

We can evaluate
\begin{align}
    B^{-1}A &= \frac{1}{\det B}\begin{pmatrix}
        b_{11} & - b_{01} \\
        - b_{10} & b_{00}
    \end{pmatrix} \begin{pmatrix}
        a_{00} & a_{01} \\
        a_{10} & a_{11}
    \end{pmatrix}  \nonumber \\
    & = \frac{1}{\det B} \begin{pmatrix}
        b_{11}a_{00}-b_{01}a_{10}  & b_{11}a_{01}-b_{01}a_{11} \\
        -b_{10}a_{00}+b_{00}a_{10}  & -b_{10}a_{01}+b_{00}a_{11}.
    \end{pmatrix} \label{eq:inversion_before}
\end{align}

Next, we assume that each value consists of the ideal value and a deviation due to some kind of noise:
\begin{align}
    n_{ij} = \tilde n_{ij} + \sigma^n_{ij}.
\end{align}
Having defined this, we look as an example at the 00 element of \autoref{eq:inversion_before}:
\begin{align}
    b_{11}a_{00}-b_{01}a_{10} &= (\tilde b_{11} + \sigma^b_{11} )(\tilde a_{00}+\sigma^a_{00}) - (\tilde b_{01} + \sigma^b_{01})(\tilde a_{10} + \sigma^a_{10}) \nonumber \\
    &= \tilde b_{11} \tilde a_{00} + \tilde a_{00}\sigma^b_{11} + b_{11}\sigma^a_{00} + \sigma^b_{11}\sigma^a_{00} -  \tilde b_{01} \tilde a_{10} - \tilde a_{10}\sigma^b_{01} - b_{01}\sigma^a_{10} - \sigma^b_{01}\sigma^a_{10} \nonumber \\
    &= \underbrace{\tilde b_{11} \tilde a_{00} - \tilde b_{01} \tilde a_{10}}_\text{value} +\underbrace{(\tilde a_{00}\sigma^b_{11} + \tilde b_{11}\sigma^a_{00} + \sigma^b_{11}\sigma^a_{00} - \tilde a_{10}\sigma^b_{01} - \tilde b_{01}\sigma^a_{10} - \sigma^b_{01}\sigma^a_{10})}_\text{deviation}
\end{align}

The solution above assumes an independent noise for each element. But in an extreme limit case of all values have the same noise, meaning
\begin{align}
    \sigma^b_{11} = \sigma^a_{00} = \sigma^b_{01} = \sigma^a_{10}= \sigma,
\end{align}
this leads to 
\begin{align}
    b_{11}a_{00}-b_{01}a_{10} &= \tilde b_{11} \tilde a_{00} - \tilde b_{01} \tilde a_{10}+(\tilde a_{00}\sigma + \tilde b_{11}\sigma + \sigma^2 - \tilde a_{10}\sigma - \tilde b_{01}\sigma - \sigma^2)\\
    &= \underbrace{\tilde b_{11} \tilde a_{00} - \tilde b_{01} \tilde a_{10}}_\text{value} +\underbrace{(\tilde a_{00}\sigma + \tilde b_{11}\sigma - \tilde a_{10}\sigma - \tilde b_{01}\sigma)}_\text{deviation}
\end{align}
where the deviations squared have cancelled out. 
Furthermore, we can write:
\begin{align}
    b_{11}a_{00}-b_{01}a_{10} &= \underbrace{\tilde b_{11} \tilde a_{00} - \tilde b_{01} \tilde a_{10}}_\text{value} +\underbrace{\sigma(\tilde a_{00}  - \tilde a_{10} + \tilde b_{11} - \tilde b_{01})}_\text{deviation},
\end{align}
which indicates smaller deviation if the matrix elements are similar. This will be system dependent. 

The same argument follows for each matrix element in $B^{-1}A$ as well as for $\det B$. Crucially, this cancellation depends on the inversion step and, thus, is an additional argument to the one in Part I.

\newpage
\section{\label{sec:add_hardware}Hardware results: Additional Figures}

\begin{figure}[htbp!]
    \centering 
    \includegraphics[width=.7\textwidth]{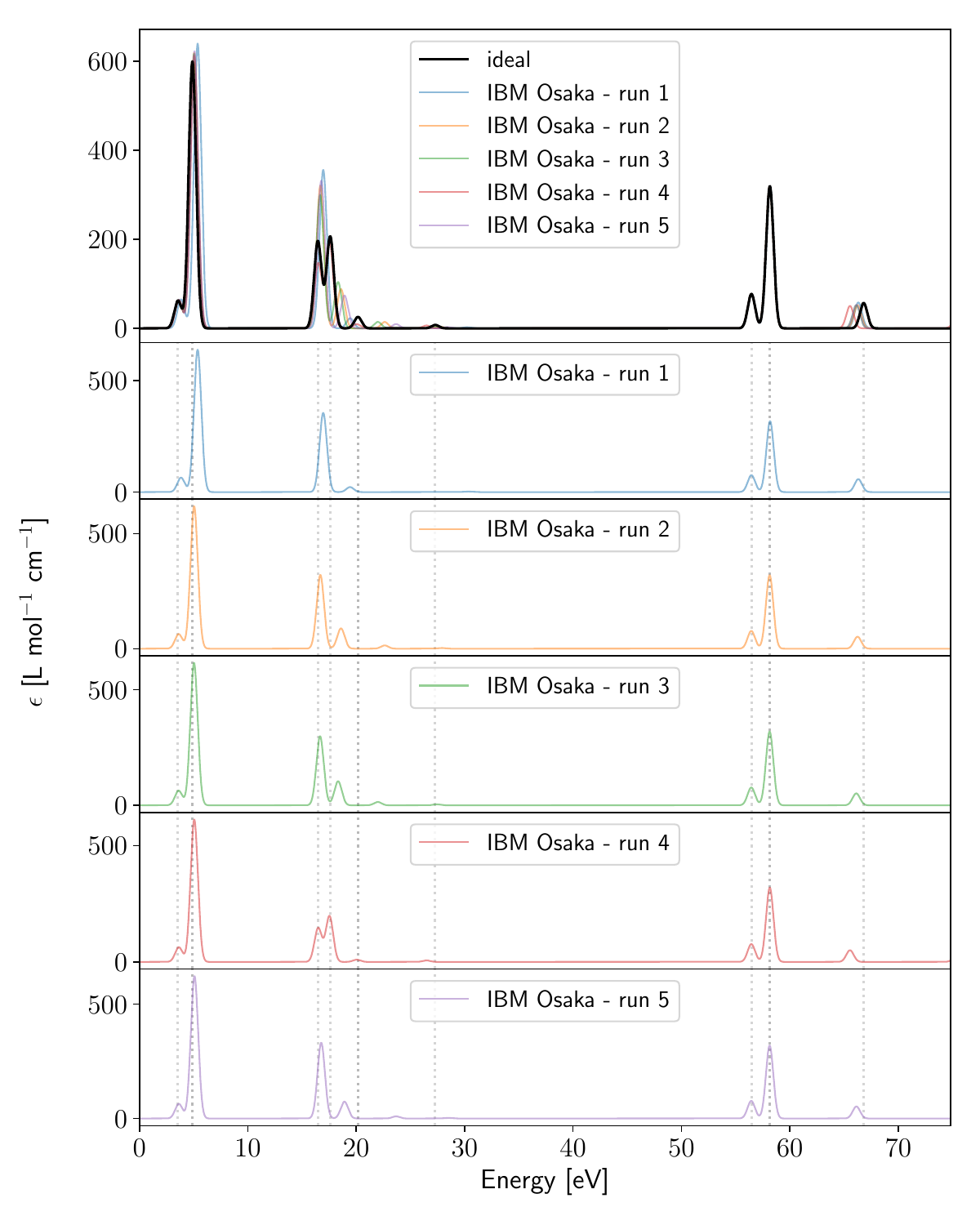}
    \caption{Absorption spectra of LiH(2,2) naive qLR run on IBM Osaka. The top panel shows in black the ideal (shot and device noise free) results and five quantum hardware runs overlaid. The lower five panels show each hardware run separately. The vertical dotted lines indicate the ideal peak positions.}
    \label{fig:LiH_harware_naive}
\end{figure}

\begin{figure}[htbp!]
    \centering 
    \includegraphics[width=.7\textwidth]{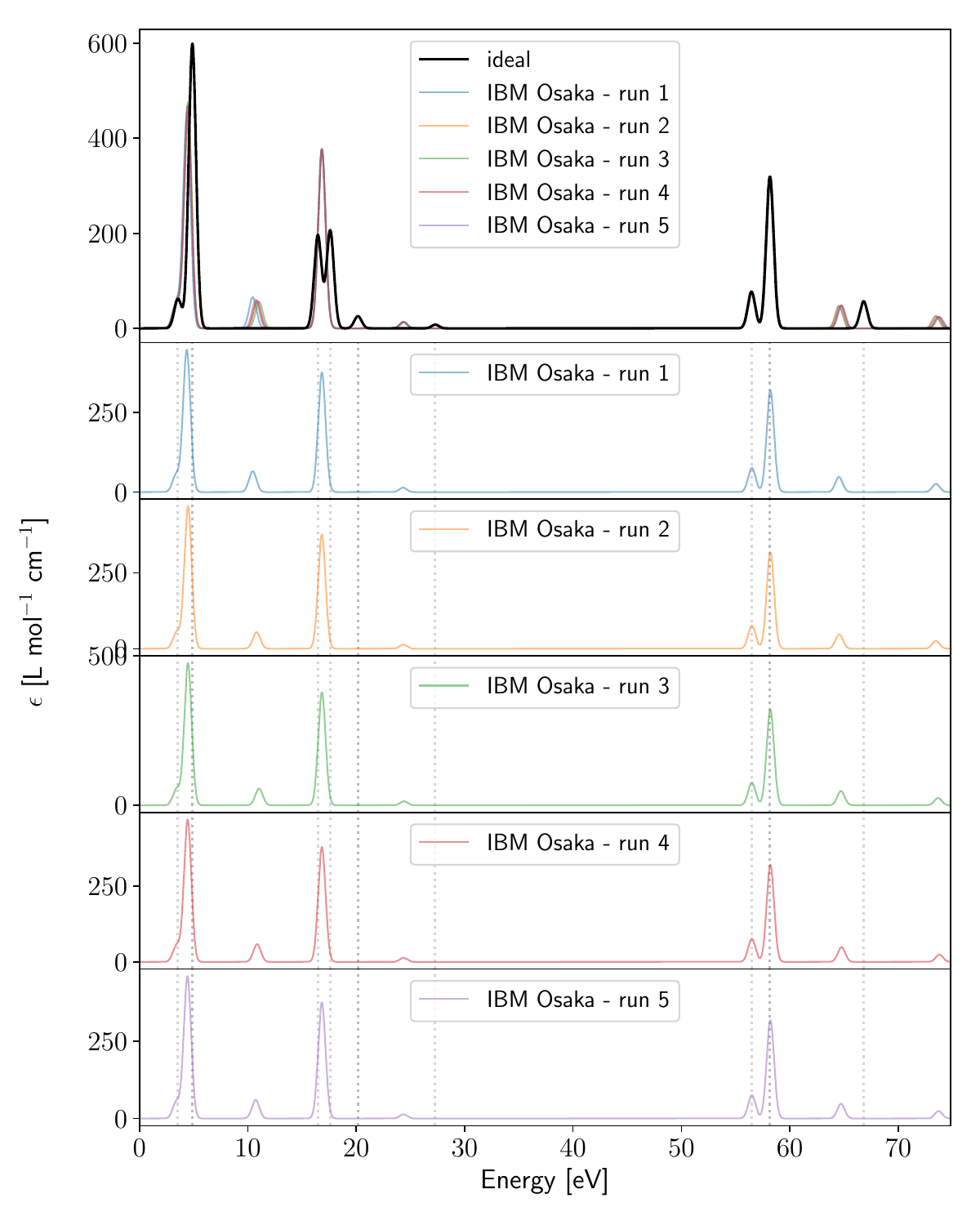}
    \caption{Absorption spectra of LiH(2,2) naive qLR run on IBM Osaka without error mitigation as described in the text \autoref{main:sec:theo:MA0}. Details as in \autoref{fig:LiH_harware_naive}}
    \label{fig:LiH_harware_naive_noM}
\end{figure}

\begin{figure}[htbp!]
    \centering 
    \includegraphics[width=.7\textwidth]{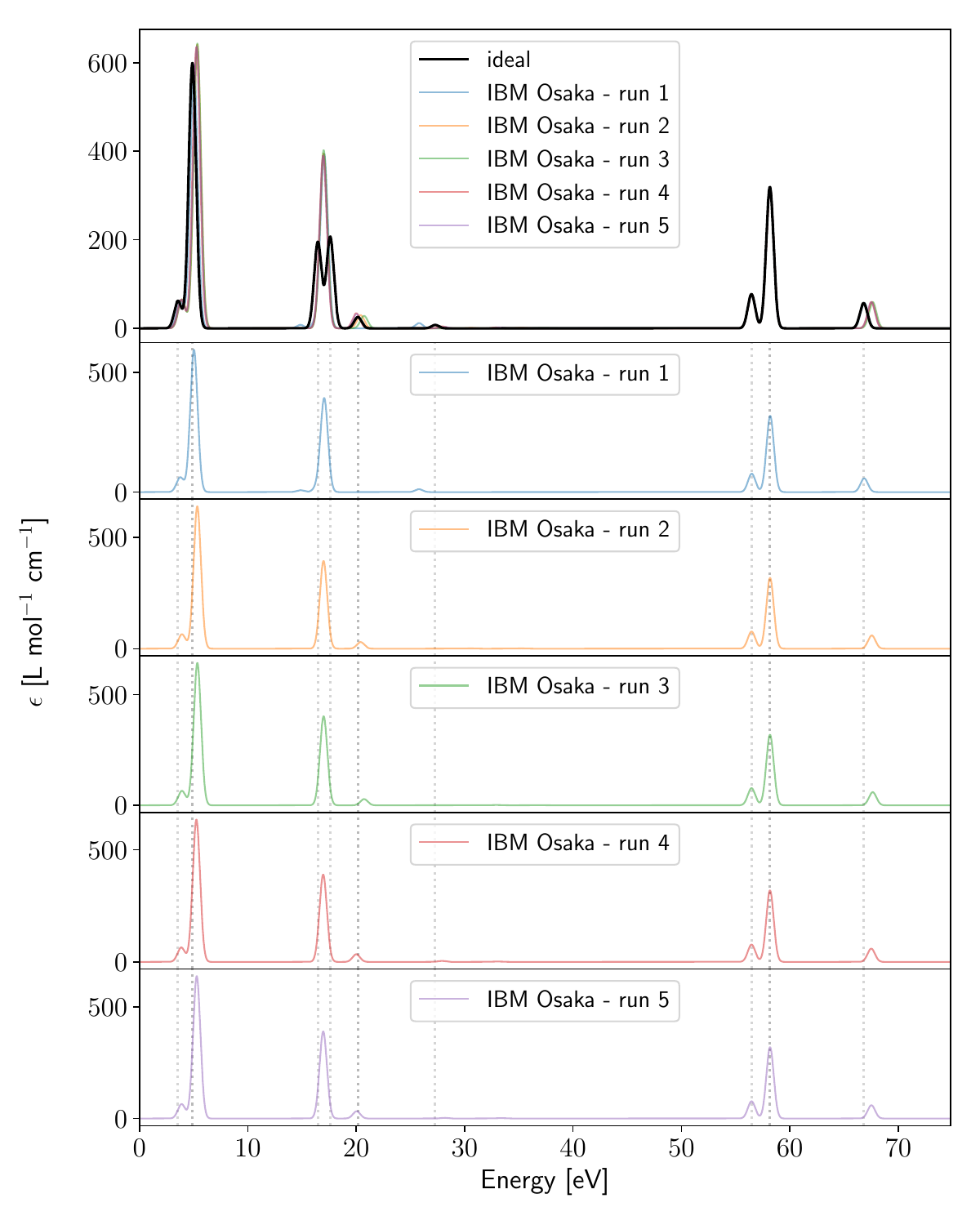}
    \caption{Absorption spectra of LiH(2,2) proj qLR run on IBM Osaka without error mitigation as described in the text \autoref{main:sec:theo:MA0}. Details as in \autoref{fig:LiH_harware_naive}}
    \label{fig:LiH_harware_proj}
\end{figure}

\begin{figure}[htbp!]
    \centering 
    \includegraphics[width=.7\textwidth]{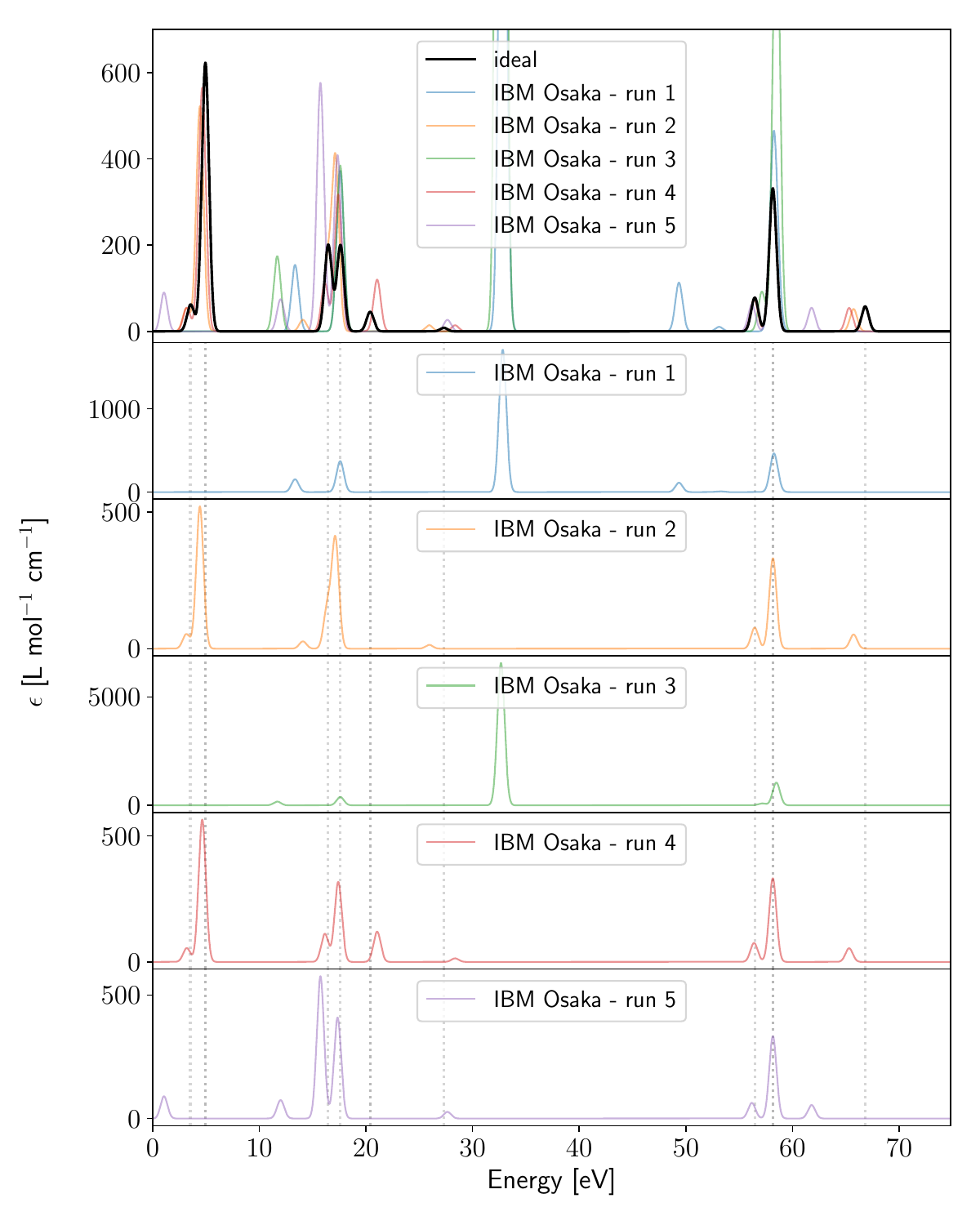}
    \caption{Absorption spectra of LiH(2,2) allproj qLR run on IBM Osaka without error mitigation as described in the text \autoref{main:sec:theo:MA0}. Details as in \autoref{fig:LiH_harware_naive}. Quantum run 1, 3 and 5 failed as they produced negative Hessian eigenvalues and were removed in post-processing (see text \autoref{main:fig:LiH_hardware}). This shows how from one consecutive run to another quantum hardware can observe sudden changes in error rate.}
    \label{fig:LiH_harware_allproj}
\end{figure}

\begin{figure}[htbp!]
    \centering 
    \includegraphics[width=.7\textwidth]{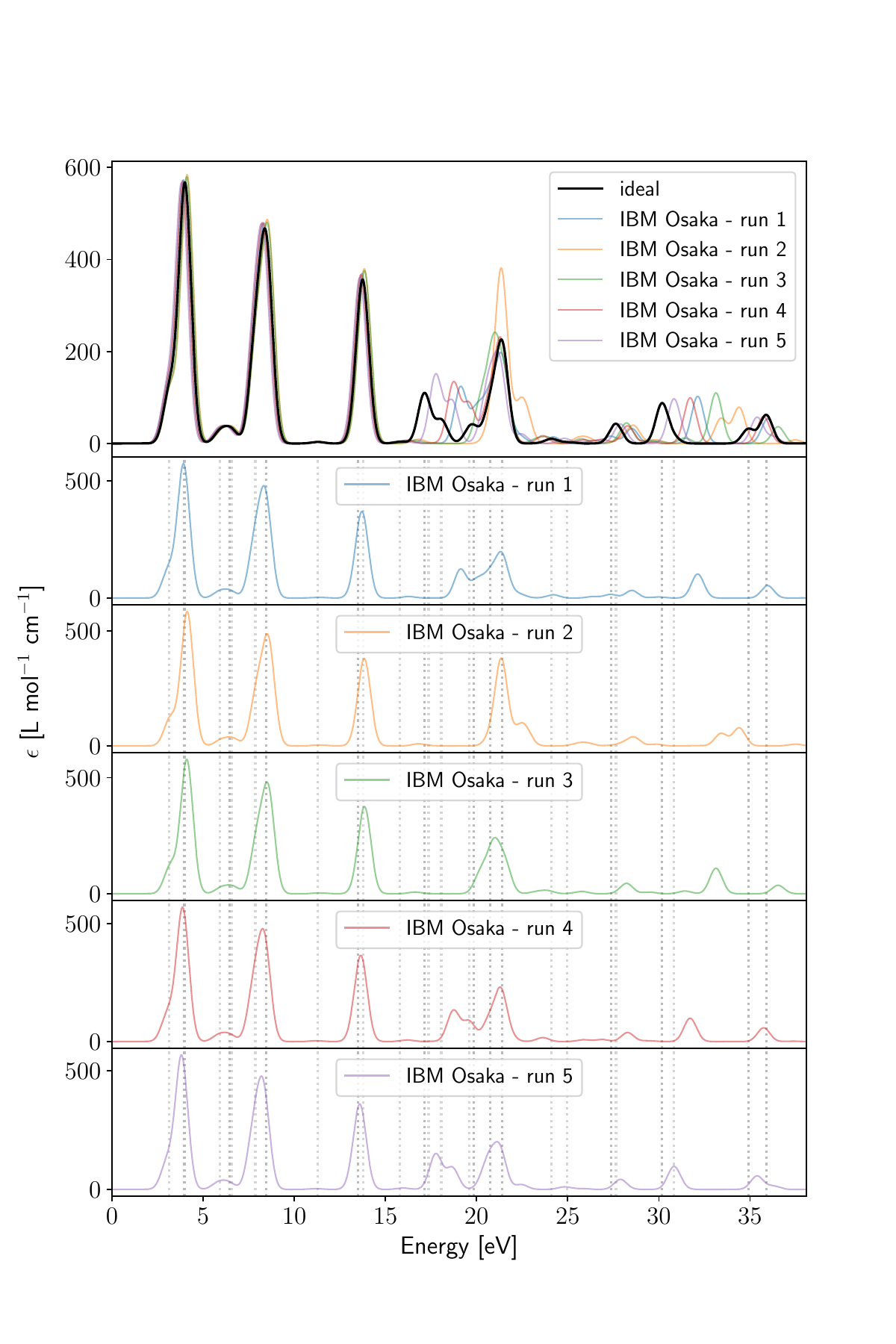}
    \caption{Low energy range absorption spectra of LiH(2,2) naive qLR run on IBM Osaka with cc-pVTZ. The top panel shows in black the ideal (shot- and 
    device-noise-free) results and five quantum hardware runs overlaid. The lower five panels show each hardware run separately. The vertical dotted lines indicate the ideal peak positions.}
    \label{fig:LiH_harware_naive_TZ_1}
\end{figure}

\begin{figure}[htbp!]
    \centering 
    \includegraphics[width=.7\textwidth]{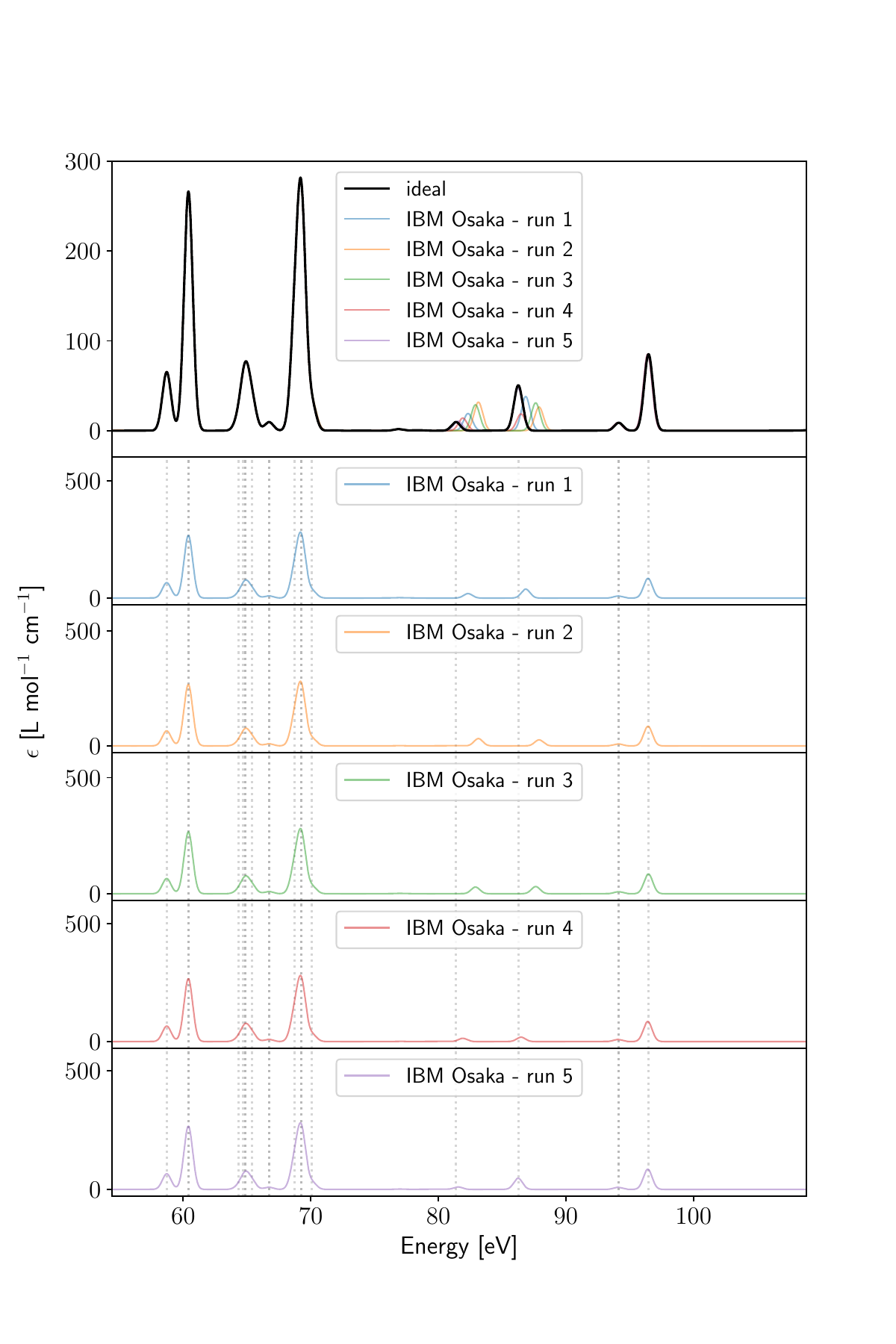}
    \caption{High energy range absorption spectra of LiH(2,2) naive qLR run on IBM Osaka with cc-pVTZ. The top panel shows in black the ideal (shot- and 
    device-noise-free) results and five quantum hardware runs overlaid. The lower five panels show each hardware run separately. The vertical dotted lines indicate the ideal peak positions.}
    \label{fig:LiH_harware_naive_TZ_2}
\end{figure}

\begin{figure}[htbp!]
    \centering 
    \includegraphics[width=\textwidth]{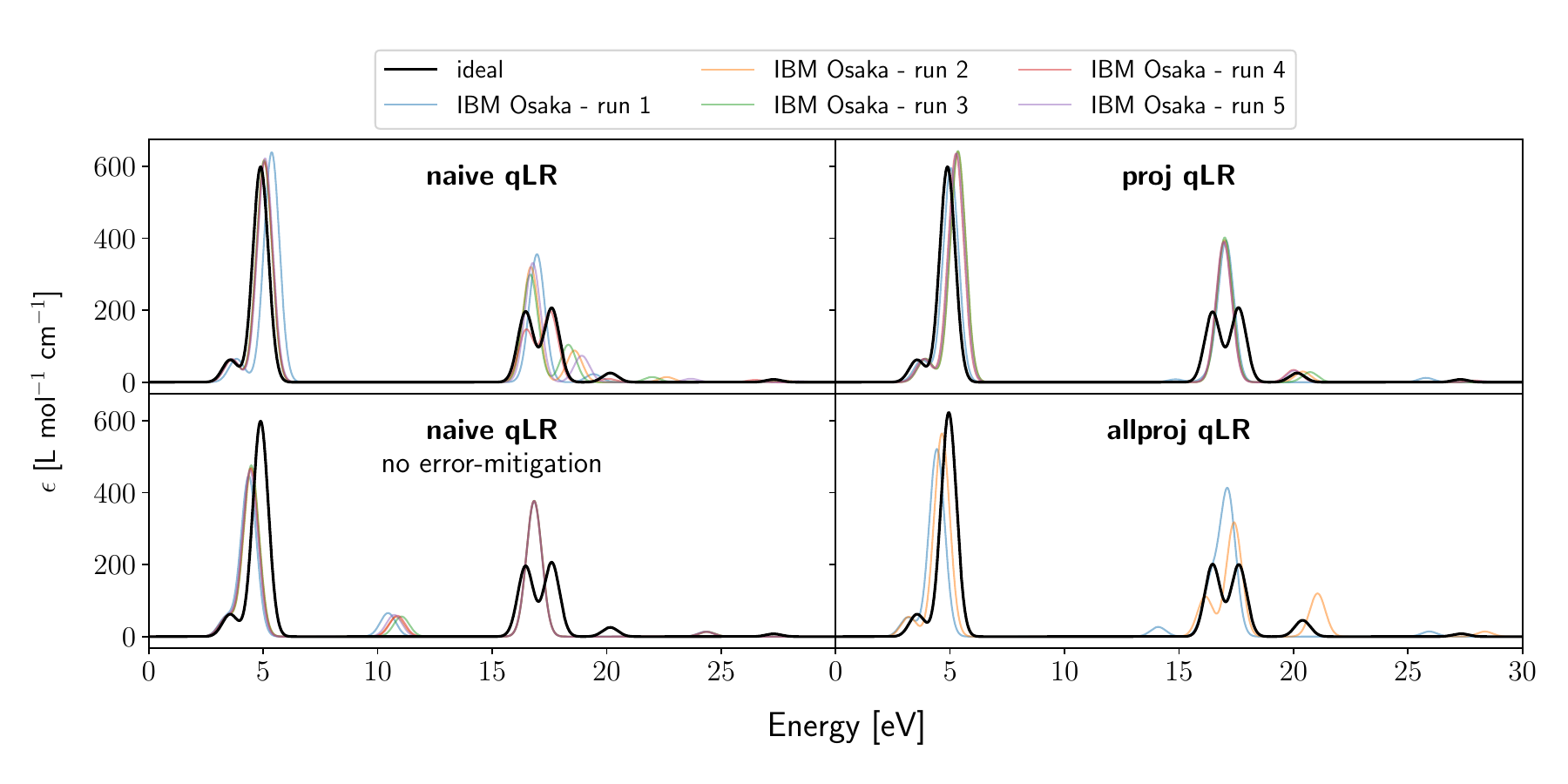}
    \caption{Zoomed in version of text \autoref{main:fig:LiH_hardware}}
    \label{fig:LiH_hardware_zoomed}
\end{figure}

\begin{figure}[htbp!]
    \centering 
    \includegraphics[width=\textwidth]{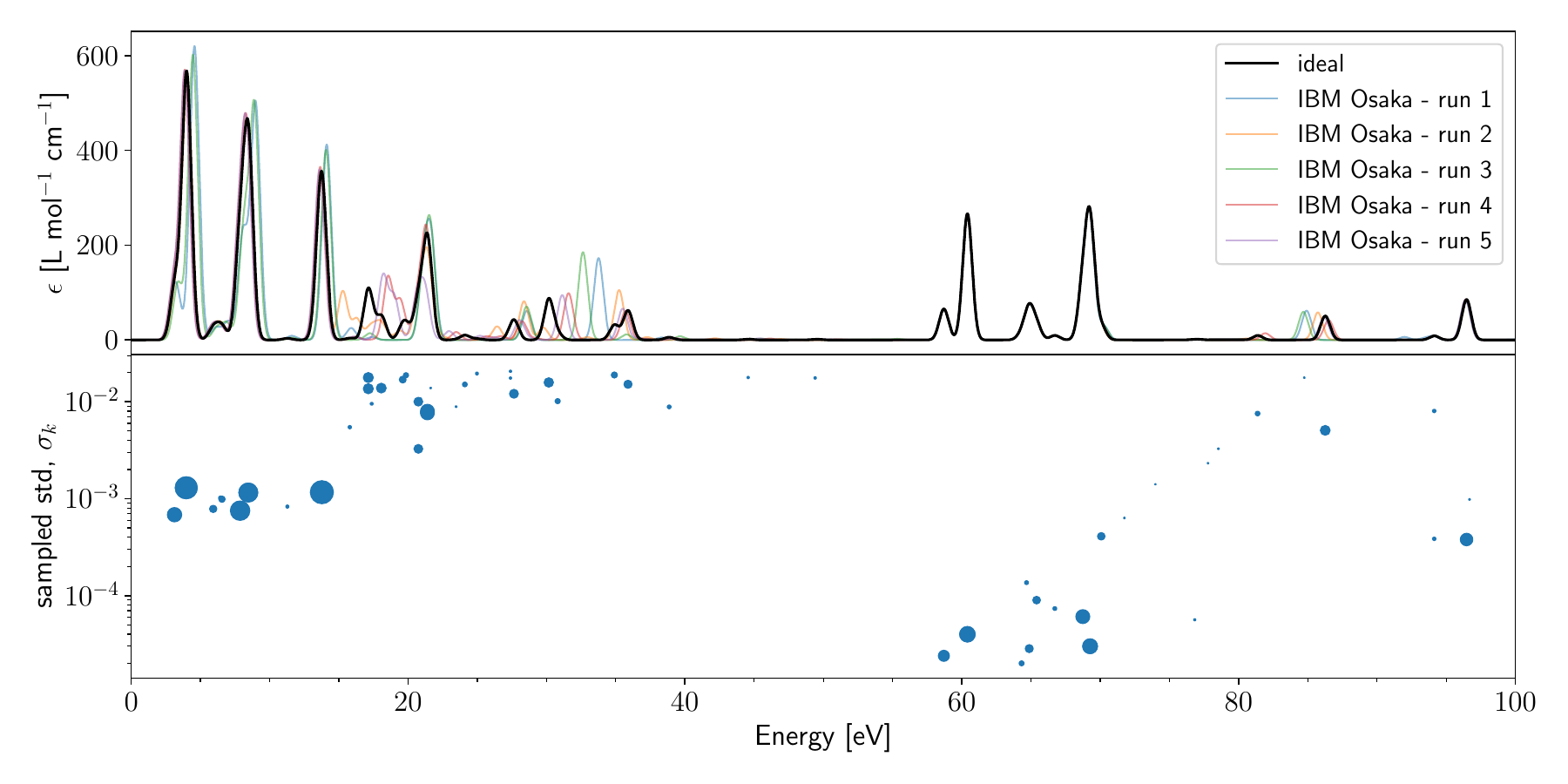}
    \caption{Top: Absorption spectra of LiH(2,2) / cc-pVTZ calculated on IBM Osaka with naive qLR using 500k shots per Pauli string. The black line indicates the ideal (shot- and 
    device-noise-free) results that coincide with classic CASSCF. Five quantum hardware runs were performed and are shown in different colours. Detailed results of each hardware run are found in \autoref{sec:add_hardware}. Bottom: Sampled standard deviation, $\sigma_k$, for 1000 samples with 100k shots on shot noise simulator. The size of the dot correlates with the oscillator strength value.}
    \label{fig:LiH_hardware_TZ_500k}
\end{figure}